\documentclass[twocolumn]{aastex62}

\def\arcmin{\hbox{$^\prime$}}
\def\arcsec{\hbox{$^{\prime\prime}$}}

\shorttitle{K2 Observations of SN 2018oh}
\shortauthors{Dimitriadis et al.}

\begin{document}

\title{K2 Observations of SN\,2018oh Reveal a Two-Component Rising Light Curve for a Type Ia Supernova}

\correspondingauthor{Georgios Dimitriadis}
\email{gdimitri@ucsc.edu}

\author{G.~Dimitriadis}
\affil{Department of Astronomy and Astrophysics, University of California, Santa Cruz, CA 95064, USA}

\author{R.~J.~Foley}
\affil{Department of Astronomy and Astrophysics, University of California, Santa Cruz, CA 95064, USA}

\author{A.~Rest}
\affiliation{Space Telescope Science Institute, 3700 San Martin Drive, Baltimore, MD 21218, USA}
\affiliation{Department of Physics and Astronomy, Johns Hopkins University, Baltimore, MD 21218, USA}

\author{D.~Kasen}
\affiliation{Department of Astronomy, University of California, Berkeley, CA 94720-3411, USA}
\affiliation{Lawrence Berkeley National Laboratory, 1 Cyclotron Road, Berkeley, California 94720, USA}

\author{A.~L.~Piro}
\affiliation{The Observatories of the Carnegie Institution for Science, 813 Santa Barbara St., Pasadena, CA 91101, USA}

\author{A.~Polin}
\affiliation{Department of Astronomy, University of California, Berkeley, CA, 94720-3411, USA}
\affiliation{Lawrence Berkeley National Laboratory, Berkeley, CA, 94720, USA}

\author{D.~O.~Jones}
\affil{Department of Astronomy and Astrophysics, University of California, Santa Cruz, CA 95064, USA}

\author{A.~Villar}
\affiliation{Harvard-Smithsonian Center for Astrophysics, 60 Garden Street, Cambridge, MA 02138, USA}

\author{G.~Narayan}
\affiliation{Department of Physics and Astronomy, Johns Hopkins University, Baltimore, MD 21218, USA}

\author{D.~ A.~Coulter}
\affil{Department of Astronomy and Astrophysics, University of California, Santa Cruz, CA 95064, USA}

\author{C.~D.~Kilpatrick}
\affil{Department of Astronomy and Astrophysics, University of California, Santa Cruz, CA 95064, USA}

\author{Y.~-C.~Pan}
\affil{Department of Astronomy and Astrophysics, University of California, Santa Cruz, CA 95064, USA}

\author{C.~Rojas-Bravo}
\affil{Department of Astronomy and Astrophysics, University of California, Santa Cruz, CA 95064, USA}

\author{O.~D.~Fox}
\affiliation{Space Telescope Science Institute, 3700 San Martin Drive, Baltimore, MD 21218, USA}

\author{S.~W.~Jha}
\affil{Department of Physics and Astronomy, Rutgers, The State University of New Jersey, 136 Frelinghuysen Road, Piscataway, NJ 08854, USA}

\author{P.~E.~Nugent}
\affiliation{Department of Astronomy, University of California, Berkeley, CA, 94720-3411, USA}
\affiliation{Lawrence Berkeley National Laboratory, Berkeley, CA, 94720, USA}

\author{A.~G.~Riess}
\affiliation{Space Telescope Science Institute, 3700 San Martin Drive, Baltimore, MD 21218, USA}
\affiliation{Department of Physics and Astronomy, Johns Hopkins University, Baltimore, MD 21218, USA}

\author{D.~Scolnic}
\affiliation{Kavli Institute for Cosmological Physics, University of Chicago, Chicago, IL 60637, USA}

\author{M.~R.~Drout}
\altaffiliation{Hubble Fellow, Dunlap Fellow}
\affiliation{The Observatories of the Carnegie Institution for Science, 813 Santa Barbara St., Pasadena, CA 91101, USA}

\nocollaboration

\collaboration{K2 Mission Team}

\author{G.~Barentsen}
\affiliation{NASA Ames Research Center, Moffett Field, CA 94035, USA}
\affiliation{Bay Area Environmental Research Institute, P.O. Box 25, Moffett Field, CA 94035, USA}

\author{J.~Dotson}
\affiliation{NASA Ames Research Center, Moffett Field, CA 94035, USA}

\author{M.~Gully-Santiago}
\affiliation{NASA Ames Research Center, Moffett Field, CA 94035, USA}
\affiliation{Bay Area Environmental Research Institute, P.O. Box 25, Moffett Field, CA 94035, USA}

\author{C.~Hedges}
\affiliation{NASA Ames Research Center, Moffett Field, CA 94035, USA}
\affiliation{Bay Area Environmental Research Institute, P.O. Box 25, Moffett Field, CA 94035, USA}

\author{A.~M.~Cody}
\affiliation{NASA Ames Research Center, Moffett Field, CA 94035, USA}
\affiliation{Bay Area Environmental Research Institute, P.O. Box 25, Moffett Field, CA 94035, USA}

\author{T.~Barclay}
\affiliation{NASA Goddard Space Flight Center, 8800 Greenbelt Rd, Greenbelt, MD 20771, USA}
\affiliation{University of Maryland, Baltimore County, 1000 Hilltop Cir, Baltimore, MD 21250, USA}

\author{S.~Howell}
\affiliation{NASA Ames Research Center, Moffett Field, CA 94035, USA}

\nocollaboration

\collaboration{KEGS}

\author{P.~Garnavich}
\affiliation{Department of Physics, University of Notre Dame, 225 Nieuwland Science Hall, Notre Dame, IN, 46556-5670, USA}

\author{B.~E.~Tucker}
\affiliation{The Research School of Astronomy and Astrophysics, Mount Stromlo Observatory, Australian National University, Canberra, ACT 2611, Australia}
\affiliation{National Centre for the Public Awareness of Science, Australian National University, Canberra, ACT 2611, Australia}
\affiliation{The ARC Centre of Excellence for All-Sky Astrophysics in 3 Dimension (ASTRO 3D), Australia}

\author{E.~Shaya}
\affiliation{Astronomy Department, University of Maryland, College Park, MD 20742-2421, USA}

\author{R.~Mushotzky}
\affiliation{Astronomy Department, University of Maryland, College Park, MD 20742-2421, USA}

\author{R.~P.~Olling}
\affiliation{Astronomy Department, University of Maryland, College Park, MD 20742-2421, USA}

\author{S.~Margheim}
\affiliation{Gemini Observatory, La Serena, Chile}

\author{A.~Zenteno}
\affiliation{Cerro Tololo Inter-American Observatory, Casilla 603, La Serena, Chile}

\nocollaboration

\collaboration{Kepler spacecraft team}

\author{J.~Coughlin}
\author{J.~E.~Van Cleve}
\affiliation{NASA Ames Research Center, Moffett Field, CA 94035, USA}
\affiliation{SETI Institute, 189 Bernardo Avenue, Mountain View, CA 94043, USA}

\author{J.~Vin\'icius de Miranda\ Cardoso}
\affiliation{NASA Ames Research Center, Moffett Field, CA 94035, USA}
\affiliation{Universidade Federal de Campina Grande, Campina Grande, Brazil}

\author{K.~A.~Larson}
\author{K.~M.~McCalmont-Everton}
\author{C.~A.~Peterson}
\author{S.~E.~Ross}
\affiliation{Ball Aerospace and Technologies Corp., Boulder, Colorado, 80301, USA}

\author{L.~H.~Reedy}
\author{D.~Osborne}
\author{C.~McGinn}
\author{L.~Kohnert}
\author{L.~Migliorini}
\author{A.~Wheaton}
\author{B.~Spencer}
\author{C.~Labonde}
\author{G.~Castillo}
\author{G.~Beerman}
\author{K.~Steward}
\author{M.~Hanley}
\author{R.~Larsen}
\author{R.~Gangopadhyay}
\author{R.~Kloetzel}
\author{T.~Weschler}
\author{V.~Nystrom}
\author{J.~Moffatt}
\author{M.~Redick}
\author{K.~Griest}
\author{M.~Packard}
\author{M.~Muszynski}
\author{J.~Kampmeier}
\author{R.~Bjella}
\author{S.~Flynn}
\author{B.~Elsaesser}
\affiliation{LASP, University of Colorado at Boulder, Boulder, CO 80303, USA}

\nocollaboration

\collaboration{Pan-STARRS}

\author{K.~C.~Chambers}
\affiliation{Institute of Astronomy, University of Hawaii, 2680 Wood-
lawn Drive, Honolulu, Hawaii 96822, USA}

\author{H.~A.~Flewelling}
\affiliation{Institute of Astronomy, University of Hawaii, 2680 Wood-
lawn Drive, Honolulu, Hawaii 96822, USA}

\author{M.~E.~Huber}
\affiliation{Institute of Astronomy, University of Hawaii, 2680 Wood-
lawn Drive, Honolulu, Hawaii 96822, USA}

\author{E.~A.~Magnier}
\affiliation{Institute of Astronomy, University of Hawaii, 2680 Wood-
lawn Drive, Honolulu, Hawaii 96822, USA}

\author{C.~Z.~Waters}
\affiliation{Institute of Astronomy, University of Hawaii, 2680 Wood-
lawn Drive, Honolulu, Hawaii 96822, USA}

\author{A.~S.~B.~Schultz}
\affiliation{Institute of Astronomy, University of Hawaii, 2680 Wood-
lawn Drive, Honolulu, Hawaii 96822, USA}

\author{J.~Bulger}
\affiliation{Institute of Astronomy, University of Hawaii, 2680 Wood-
lawn Drive, Honolulu, Hawaii 96822, USA}

\author{T.~B.~Lowe}
\affiliation{Institute of Astronomy, University of Hawaii, 2680 Wood-
lawn Drive, Honolulu, Hawaii 96822, USA}

\author{M.~Willman}
\affiliation{Institute of Astronomy, University of Hawaii, 2680 Wood-
lawn Drive, Honolulu, Hawaii 96822, USA}

\author{S.~J.~Smartt}
\affiliation{Astrophysics Research Centre, School of Mathematics and Physics, Queens University Belfast, Belfast BT7 1NN, UK}

\author{K.~W.~Smith}
\affiliation{Astrophysics Research Centre, School of Mathematics and Physics, Queens University Belfast, Belfast BT7 1NN, UK}

\nocollaboration

\collaboration{DECam}

\author{S.~Points}
\affiliation{Cerro Tololo Inter-American Observatory, Casilla 603, La Serena, Chile}

\author{G.~M.~Strampelli}
\affiliation{Space Telescope Science Institute, 3700 San Martin Drive, Baltimore, MD 21218, USA}
\affiliation{University of La Laguna,  Calle Padre Herrera, 38200 San Cristóbal de La Laguna, Santa Cruz de Tenerife, Spain}

\nocollaboration

\collaboration{ASAS-SN}

\author{J.~Brimacombe}
\affiliation{Coral Towers Observatory, Cairns, Queensland 4870, Australia}

\author{P.~Chen}
\affiliation{Kavli Institute for Astronomy and Astrophysics, Peking University, Yi He Yuan Road 5, Hai Dian District, Beijing 100871, China}

\author{J.~A.~Mu\~{n}oz}
\affiliation{Departamento de Astronom\'{\i}a y Astrof\'{\i}sica, Universidad de Valencia, E-46100 Burjassot, Valencia, Spain}
\affiliation{Observatorio Astron\'omico, Universidad de Valencia, E-46980 Paterna, Valencia, Spain}

\author{R.~L.~Mutel}
\affiliation{Department of Physics and Astronomy, University of Iowa, Iowa City, IA 52242, USA}

\author{J.~Shields}
\affiliation{Department of Astronomy, The Ohio State University, 140 West 18th Avenue, Columbus, OH 43210, USA}

\author{P.~J.~Vallely}
\affiliation{Department of Astronomy, The Ohio State University, 140 West 18th Avenue, Columbus, OH 43210, USA}

\author{S.~Villanueva~Jr.}
\affiliation{Department of Astronomy, The Ohio State University, 140 West 18th Avenue, Columbus, OH 43210, USA}

\nocollaboration

\collaboration{PTSS/TNTS}

\author{W.~Li}
\affil{Physics Department and Tsinghua Center for Astrophysics (THCA), Tsinghua University, Beijing, 100084, China}
\affiliation{Las Cumbres Observatory, 6740 Cortona Dr Ste 102, Goleta, CA 93117-5575, USA}
\author{X.~Wang}
\affil{Physics Department and Tsinghua Center for Astrophysics (THCA), Tsinghua University, Beijing, 100084, China}
\author{J.~Zhang}
\affil{Yunnan Astronomical Observatory of China, Chinese Academy of Sciences, Kunming, 650011, China}
\affil{Key Laboratory for the Structure and Evolution of Celestial Objects, Chinese Academy of Sciences, Kunming 650216, China}
\affil{Center for Astronomical Mega-Science, Chinese Academy of Sciences, 20A Datun Road, Chaoyang District, Beijing, 100012, China}
\author{H.~Lin}
\affil{Physics Department and Tsinghua Center for Astrophysics (THCA), Tsinghua University, Beijing, 100084, China}
\author{J.~Mo}
\affil{Physics Department and Tsinghua Center for Astrophysics (THCA), Tsinghua University, Beijing, 100084, China}
\author{X.~Zhao}
\affil{School of Science, Tianjin University of Technology, Tianjin, 300384, China}
\author{H.~Sai}
\affil{Physics Department and Tsinghua Center for Astrophysics (THCA), Tsinghua University, Beijing, 100084, China}
\author{X.~Zhang}
\affil{Physics Department and Tsinghua Center for Astrophysics (THCA), Tsinghua University, Beijing, 100084, China}
\author{K.~Zhang}
\affil{Physics Department and Tsinghua Center for Astrophysics (THCA), Tsinghua University, Beijing, 100084, China}
\affil{Department of Astronomy, University of Texas at Austin, Austin, TX, 78712, USA}
\author{T.~Zhang}
\affil{National Astronomical Observatory of China, Chinese Academy of Sciences, Beijing, 100012, China}
\author{L.~Wang}
\affil{National Astronomical Observatory of China, Chinese Academy of Sciences, Beijing, 100012, China}
\affil{Chinese Academy of Sciences South America Center for Astronomy, China-Chile Joint Center for Astronomy, Camino El Observatorio 1515, Las Condes, Santiago, Chile}
\author{J.~Zhang}
\affil{Physics Department and Tsinghua Center for Astrophysics (THCA), Tsinghua University, Beijing, 100084, China}
\author{E.~Baron}
\affil{Homer L. Dodge Department of Physics and Astronomy, University of Oklahoma, Norman, OK}
\author{J.~M.~DerKacy}
\affil{Homer L. Dodge Department of Physics and Astronomy, University of Oklahoma, Norman, OK}
\author{L.~Li}
\affil{Physics Department and Tsinghua Center for Astrophysics (THCA), Tsinghua University, Beijing, 100084, China}
\author{Z.~Chen}
\affil{Physics Department and Tsinghua Center for Astrophysics (THCA), Tsinghua University, Beijing, 100084, China}
\author{D.~Xiang}
\affil{Physics Department and Tsinghua Center for Astrophysics (THCA), Tsinghua University, Beijing, 100084, China}
\author{L.~Rui}
\affil{Physics Department and Tsinghua Center for Astrophysics (THCA), Tsinghua University, Beijing, 100084, China}
\author{L.~Wang}
\affil{Purple Mountain Observatory, Chinese Academy of Sciences, Nanjing 210034, China}
\affil{George P. and Cynthia Woods Mitchell Institute for Fundamental Physics $\&$ Astronomy, Texas A. $\&$ M.
University, Department of Physics and Astronomy, 4242 TAMU, College Station, TX 77843, USA}
\author{F.~Huang}
\affil{Department of Astronomy, School of Physics and Astronomy, Shanghai Jiao Tong University, Shanghai, 200240, China}
\affil{Physics Department and Tsinghua Center for Astrophysics (THCA), Tsinghua University, Beijing, 100084, China}
\author{X.~Li}
\affil{Physics Department and Tsinghua Center for Astrophysics (THCA), Tsinghua University, Beijing, 100084, China}

\nocollaboration

\collaboration{Las Cumbres Observatory}

\author{G.~Hosseinzadeh}
\affiliation{Las Cumbres Observatory, 6740 Cortona Dr Ste 102, Goleta, CA 93117-5575, USA}
\affiliation{Department of Physics, University of California, Santa Barbara, CA 93106-9530, USA}

\author{D.~A.~Howell}
\affiliation{Las Cumbres Observatory, 6740 Cortona Dr Ste 102, Goleta, CA 93117-5575, USA}
\affiliation{Department of Physics, University of California, Santa Barbara, CA 93106-9530, USA}

\author{I.~Arcavi}
\altaffiliation{Einstein Fellow}
\affiliation{Las Cumbres Observatory, 6740 Cortona Dr Ste 102, Goleta, CA 93117-5575, USA}
\affiliation{Department of Physics, University of California, Santa Barbara, CA 93106-9530, USA}
\affiliation{The Raymond and Beverly Sackler School of Physics and Astronomy, Tel Aviv University, Tel Aviv 69978, Israel}

\author{D.~Hiramatsu}
\affiliation{Las Cumbres Observatory, 6740 Cortona Dr Ste 102, Goleta, CA 93117-5575, USA}
\affiliation{Department of Physics, University of California, Santa Barbara, CA 93106-9530, USA}

\author{J.~Burke}
\affiliation{Las Cumbres Observatory, 6740 Cortona Dr Ste 102, Goleta, CA 93117-5575, USA}
\affiliation{Department of Physics, University of California, Santa Barbara, CA 93106-9530, USA}

\author{S.~Valenti}
\affiliation{Department of Physics, University of California, 1 Shields Ave, Davis, CA 95616-5270, USA}

\nocollaboration

\collaboration{ATLAS}

\author{J.~L.~Tonry}
\affiliation{Institute of Astronomy, University of Hawaii, 2680 Wood-
lawn Drive, Honolulu, Hawaii 96822, USA}

\author{L.~Denneau}
\affiliation{Institute of Astronomy, University of Hawaii, 2680 Wood-
lawn Drive, Honolulu, Hawaii 96822, USA}

\author{A.~N.~Heinze}
\affiliation{Institute of Astronomy, University of Hawaii, 2680 Wood-
lawn Drive, Honolulu, Hawaii 96822, USA}

\author{H.~Weiland}
\affiliation{Institute of Astronomy, University of Hawaii, 2680 Wood-
lawn Drive, Honolulu, Hawaii 96822, USA}

\author{B.~Stalder}
\affiliation{LSST, 950 North Cherry Avenue, Tucson, AZ 85719, USA}

\nocollaboration

\collaboration{Konkoly}

\author{J.~Vink\'o}
\affiliation{Konkoly Observatory, MTA CSFK, Konkoly Thege M. ut 15-17, Budapest, 1121, Hungary}
\affiliation{Department of Optics \& Quantum Electronics, University of Szeged, Dom ter 9, Szeged, 6720 Hungary}
\affiliation{Department of Astronomy, University of Texas at Austin, Austin, TX, 78712, USA}

\author{K.~S\'arneczky}
\affiliation{Konkoly Observatory, MTA CSFK, Konkoly Thege M. ut 15-17, Budapest, 1121, Hungary}

\author{A.~P\'al}
\affiliation{Konkoly Observatory, MTA CSFK, Konkoly Thege M. ut 15-17, Budapest, 1121, Hungary}

\author{A.~B\'odi},
\affiliation{Konkoly Observatory, MTA CSFK, Konkoly Thege M. ut 15-17, Budapest, 1121, Hungary}
\affiliation{MTA CSFK Lend\"ulet Near-Field Cosmology Research Group}

\author{Zs.~Bogn\'ar}
\affiliation{Konkoly Observatory, MTA CSFK, Konkoly Thege M. ut 15-17, Budapest, 1121, Hungary}

\author{B.~Cs\'ak}
\affiliation{Konkoly Observatory, MTA CSFK, Konkoly Thege M. ut 15-17, Budapest, 1121, Hungary}

\author{B.~Cseh}
\affiliation{Konkoly Observatory, MTA CSFK, Konkoly Thege M. ut 15-17, Budapest, 1121, Hungary}

\author{G.~Cs\"ornyei}
\affiliation{Konkoly Observatory, MTA CSFK, Konkoly Thege M. ut 15-17, Budapest, 1121, Hungary}

\author{O.~Hanyecz}
\affiliation{Konkoly Observatory, MTA CSFK, Konkoly Thege M. ut 15-17, Budapest, 1121, Hungary}

\author{B.~Ign\'acz}
\affiliation{Konkoly Observatory, MTA CSFK, Konkoly Thege M. ut 15-17, Budapest, 1121, Hungary}

\author{Cs.~Kalup}
\affiliation{Konkoly Observatory, MTA CSFK, Konkoly Thege M. ut 15-17, Budapest, 1121, Hungary}

\author{R.~K\"onyves-T\'oth}
\affiliation{Konkoly Observatory, MTA CSFK, Konkoly Thege M. ut 15-17, Budapest, 1121, Hungary}

\author{L.~Kriskovics}
\affiliation{Konkoly Observatory, MTA CSFK, Konkoly Thege M. ut 15-17, Budapest, 1121, Hungary}

\author{A.~Ordasi}
\affiliation{Konkoly Observatory, MTA CSFK, Konkoly Thege M. ut 15-17, Budapest, 1121, Hungary}

\author{I.~Rajmon}
\affiliation{Berzsenyi D\'aniel High School, K\'arp\'at utca 49-53, Budapest, 1133, Hungary}

\author{A.~S\'odor}
\affiliation{Konkoly Observatory, MTA CSFK, Konkoly Thege M. ut 15-17, Budapest, 1121, Hungary}

\author{R.~Szab\'o}
\affiliation{Konkoly Observatory, MTA CSFK, Konkoly Thege M. ut 15-17, Budapest, 1121, Hungary}

\author{R.~Szak\'ats}
\affiliation{Konkoly Observatory, MTA CSFK, Konkoly Thege M. ut 15-17, Budapest, 1121, Hungary}

\author{G.~Zsidi}
\affiliation{Konkoly Observatory, MTA CSFK, Konkoly Thege M. ut 15-17, Budapest, 1121, Hungary}

\nocollaboration

\collaboration{ePESSTO}

\author{S.~C.~Williams}
\affiliation{Department of Physics, Lancaster University, Lancaster, LA1 4YB, UK}

\author{J.~Nordin}
\affiliation{Institute of Physics, Humboldt-Universit\"at zu Berlin, Newtonstr. 15, 12489 Berlin, Germany}

\author{R.~Cartier}
\affiliation{Cerro Tololo Inter-American Observatory, Casilla 603, La Serena, Chile}

\author{C.~Frohmaier}
\affiliation{Institute of Cosmology and Gravitation, University of Portsmouth, Portsmouth, PO1 3FX, UK}

\author{L.~Galbany}
\affiliation{PITT PACC, Department of Physics and Astronomy, University of Pittsburgh, Pittsburgh, PA 15260, USA}

\author{C.~P.~Guti\'errez}
\affiliation{Department of Physics and Astronomy, University of Southampton, Southampton, SO17 1BJ, UK}

\author{I.~Hook}
\affiliation{Department of Physics, Lancaster University, Lancaster, LA1 4YB, UK}

\author{C.~Inserra}
\affiliation{Department of Physics and Astronomy, University of Southampton, Southampton, SO17 1BJ, UK}

\author{M.~Smith}
\affiliation{Department of Physics and Astronomy, University of Southampton, Southampton, SO17 1BJ, UK}

\nocollaboration

\collaboration{University of Arizona}

\author{D.~J.~Sand}
\affil{Steward Observatory, University of Arizona, 933 North Cherry Avenue, Rm. N204, Tucson, AZ 85721-0065, USA}

\author{J.~E.~Andrews}
\affil{Steward Observatory, University of Arizona, 933 North Cherry Avenue, Rm. N204, Tucson, AZ 85721-0065, USA}

\author{N.~Smith}
\affil{Steward Observatory, University of Arizona, 933 North Cherry Avenue, Rm. N204, Tucson, AZ 85721-0065, USA}

\author{C.~Bilinski}
\affil{Steward Observatory, University of Arizona, 933 North Cherry Avenue, Rm. N204, Tucson, AZ 85721-0065, USA}

\begin{abstract}
We present an exquisite, 30-min cadence \textit{Kepler} (K2) light curve of the Type Ia supernova (SN~Ia) 2018oh (ASASSN-18bt), starting weeks before explosion, covering the moment of explosion and the subsequent rise, and continuing past peak brightness.  These data are supplemented by multi-color Pan-STARRS1 and CTIO 4-m DECam observations obtained within hours of explosion.  The K2 light curve has an unusual two-component shape, where the flux rises with a steep linear gradient for the first few days, followed by a quadratic rise as seen for typical SNe~Ia.  This ``flux excess'' relative to canonical SN~Ia behavior is confirmed in our $i$-band light curve, and furthermore, SN\,2018oh is especially blue during the early epochs. The flux excess peaks 2.14$\pm0.04$~days after explosion, has a FWHM of 3.12$\pm0.04$~days, a blackbody temperature of $T=17,500^{+11,500}_{-9,000}$\,K, a peak luminosity of $4.3\pm0.2\times10^{37}\,{\rm erg\,s^{-1}}$, and a total integrated energy of $1.27\pm0.01\times10^{43}\,{\rm erg}$.  We compare SN\,2018oh to several models that may provide additional heating at early times, including collision with a companion and a shallow concentration of radioactive nickel. While all of these models generally reproduce the early K2 light curve shape, we slightly favor a companion interaction, at a distance of $\sim${}$2\times10^{12}\,{\rm cm}$ based on our early color measurements, although the exact distance depends on the uncertain viewing angle. Additional confirmation of a companion interaction in future modeling and observations of SN\,2018oh would  provide strong support for a single-degenerate progenitor system.
\end{abstract}


\section{Introduction} \label{sec:intro}

Through a combination of theoretical arguments and strong observational
constraints, it has long been understood that Type Ia supernovae (SNe~Ia) are the result
of a thermonuclear explosion of a carbon/oxygen white dwarf (WD)
\citep[e.g.,][]{Hoyle60, Colgate69, Woosley86, Bloom12} in a binary
system.  Nevertheless, despite SNe~Ia being used to discover the accelerating
expansion of the Universe two decades ago \citep{Riess98AJ,Perlmutter99}
and continuing to be a powerful dark energy probe
\citep[e.g.,][]{Scolnic18,Jones18}, we still do not know the nature
of their progenitor systems, whether they come from multiple
progenitor scenarios, and if so, in what proportion.

Roughly speaking, possible SN progenitor systems can be separated into two
main classes (or channels): the single-degenerate (SD) channel, where the
primary WD accretes material from a non-degenerate companion
triggering a thermonuclear runaway near the Chandrasekhar mass
($\mathrm{M_{Ch}}$) \citep[e.g.,][]{Whelan73ApJ}, and the
double-degenerate (DD) channel where the SN is triggered by the merger
of two WDs \citep[e.g.;][]{Iben84ApJS}. Numerical modeling of
explosions \citep[e.g.,][]{Hillebrandt00ARAA,Hillebrandt13FrPhy} combined
with radiative hydrodynamic
modeling \citep[e.g.,][]{Kasen09Natur,Woosley11ApJ,Pakmor12ApJ,Sim12MNRAS,Sim13MNRAS}
indicate that the basic properties of the SN~Ia population can be reproduced
by either scenario.  Therefore, we must turn to observations to further
constrain the possible progenitor systems of SNe~Ia.

Thus far, the observations have been similarly limited, and are often
inconsistent with a single scenario.  No SN~Ia progenitor system has
yet been directly observed in the handful of SNe~Ia with reasonably
deep pre-explosion images \citep{Li11Natur, Goobar14ApJ, Kelly14ApJ} (although one has for a peculiar WD SN, the SN~Iax~2012Z; \citealt{McCully14Natur}). However, the images were not sufficiently deep to exclude all SD progenitor systems.  On a different approach, a search for the surviving non-degenerate companion star at the central regions of SN remnants (SNRs), believed to have a SN Ia origin, also excludes WD + sub-giant or red giant (RG) systems \citep{Kerzendorf12ApJ...759....7K,Schaefer12Natur.481..164S,Kerzendorf14ApJ...782...27K}.  Nonetheless, several indirect observations can reveal the nature of the companion with some scenarios having specific and distinct observational predictions.

Observing SNe~Ia as close to explosion as possible can provide
unique information for distinguishing between progenitor scenarios. For example,
the earliest moments can be dominated by the shock cooling of the exploding
WD \citep{Piro10ApJ}, which was used in the case of SN 2011fe to constrain the explosion
to be coming from a degenerate star \citep{Bloom12}. For SD progenitor systems containing a Roche-lobe-filling companion, signs of the SN ejecta interacting with the non-degenerate companion star are expected for some lines of sight \citep{Kasen10ApJ}. This
produces strong X-ray and UV/optical emission that will
surpass the radioactive luminosity of the SN at these early epochs.
The amount of observed flux depends on the
viewing angle and the distance between the exploding WD and the
companion --- which given the Roche-lobe overflow assumption, provides
constraints on the companion star radius.  Specifically, evolved red
giants are expected to produce more flux than smaller stars.

While early excess emission is a robust prediction for the
Roche-lobe-filling SD scenarios, other physical phenomena can possibly
also produce early heating.  In particular,
$^{56}$Ni near the surface (i.e., with a mass fraction
exceeding that of lower layers) should also introduce flux in excess
of the canonical ``expanding fireball'' model \citep{Piro13ApJ}.  This
scenario, which can occur for both progenitor channels, can conceal
or resemble interaction models.  A specific explosion model that can
produce such a configuration is the double-detonation explosion of a
sub-$\mathrm{M_{ch}}$ WD, where the detonation of a surface helium
layer will produce significant surface $^{56}$Ni
\citep{Noebauer17MNRAS}.  \citet{Piro16ApJ} also found that shallow
$^{56}$Ni distributions and/or interaction with circumstellar material
(CSM) expelled during a DD merger can modulate the early light curve
shape.

Observations early enough and with sufficient cadence to search for
these early light curve features are still relatively rare.  Nearby events, such as SNe\,2011fe \citep{Nugent11,Bloom12}, 2014J \citep{Goobar14ApJ} and ASASSN-14lp \citep{Shappee2016ApJ} provide upper limits to the potential
separation distance of the companion, ruling out stars more evolved
than a RG, while for SN\,2009ig, a small blue excess is
attributed to the unusual color evolution of the particular event
\citep{Foley12ApJ}.  SN~Ia sample studies \citep{Hayden10ApJ2, Bianco11ApJ,Tucker11ApSS, Ganeshalingam11MNRAS, Brown12ApJ} exclude RGs for a fraction of the events, allowing less-evolved stars as companions.

On the other hand, two SNe, SNe\,2012cg and 2017cbv, have early light
curves that are somewhat consistent with interaction with a companion
star.  \citet{Marion16ApJ} finds that interaction with a
6~$\mathrm{M_{\odot}}$ main sequence (MS) star can explain the early
UV/optical excess of SN\,2012cg. For SN\,2017cbv, \citet{Hosseinzadeh17ApJ}, analyzing many possibilities, favors an interaction with a subgiant companion, within the uncertainties in the modeling. Both of these
interpretations have been questioned by \citet{Shappee18ApJ} and
\citet{Sand18arXiv}, respectively, where the authors disfavor a
non-degenerate companion, based, in part, on non-detections of
stripped hydrogen or helium (within some limits) in nebular spectra.  For SN\,2012fr, \citealt{Contreras18ApJ...859...24C} find an initial slow, nearly linear rise in luminosity, followed by a faster rising phase, and attribute it to a moderate amount of $^{56}$Ni mixing in the ejecta, while for the almost-linear rise of iPTF\,16abc, \citealt{Miller18ApJ} argue in favor of either ejecta-mixing or pulsational delayed-detonation models.  In the case of the SN\,2002es-like \citep{Ganeshalingam12ApJ} iPTF\,14atg \citep{Cao16ApJ},
data are compatible with a companion at a separation of
70~$\mathrm{R_{\odot}}$, with \citealt{Kromer16MNRAS}, using numerical
simulations of explosion models, finding difficulties reconciling its
peculiar spectral evolution with a non-degenerate companion. Finally, \citealt{Jiang17Natur} show an early red flux excess for MUSSES1604D and, comparing different scenarios, favor a double detonation.

To search for such companion-shock emission, one would ideally conduct
a survey with continuous, high-cadence observations to precisely
constrain the explosion time and either track or constrain any
possible early-time excess flux.  The \textit{Kepler} telescope
\citep{Haas10ApJ} with its wide field of view and 30-minute cadence,
continuous observations is particularly well suited to discover SNe
within moments of explosion and continuously monitor those SNe
\citep[for recent transient studies with \textit{Kepler}
see][]{Garnavich16ApJ, Rest18NatAs}.  \textit{Kepler} has the ability
to observe thousands of galaxies at a time and therefore has the
potential to discover $\sim$10 SNe a month if the observations are
devoted to relatively nearby galaxies.  During the main
\textit{Kepler} mission, \citet{Olling15Natur} discovered 3 likely
SNe~Ia with extraordinary coverage from the moments of explosion
through the rise and decline of the SNe.  Despite these
extraordinary observations, there was no significant detection of
interaction.

Nevertheless, it has been demonstrated that \textit{Kepler} has unique
capabilities for precise monitoring of the earliest phases after a SN
explosion.  To this end, the successor of the \textit{Kepler} mission,
K2, has dedicated a substantial number of targets during Campaign 16,
lasting from December 7, 2017 to February 25, 2018, to the K2
Supernova Cosmology Experiment (K2 SCE).  Significant advantages of K2
SCE over previous \textit{Kepler} mission SN studies include (1)
monitoring about 50 times as many galaxies (although for a shorter
time) and (2) being ``forward-facing,'' where the field is pointed
roughly away from the Sun, allowing for simultaneous ground-based
observations of all transients discovered in the Campaign 16 field.

In this paper, we present observations of SN\,2018oh, a normal SN~Ia
whose host galaxy was monitored by the K2 SCE starting before
explosion, continuing to first light, and through peak brightness.  In
addition to its impressive K2 light curve, SN\,2018oh SN was
extensively monitored by many ground-based facilities.  In this paper,
we focus on the first week after explosion.  In the data, we robustly
identify, with unprecedented photometric coverage, an excess early time rise component.

This work is part of a series of papers analyzing SN\,2018oh: Shappee et~al.\ (2018b) provide an alternative analysis of the K2 light curve data of the SN and Li et~al.\ (2018) present the photometric and spectroscopic properties of the SN near and after- peak brightness. 

This paper is organized as follows: In Section~\ref{sec:disc_early_phot}, we present the discovery of SN\,2018oh and the early-time data we use in this paper, including the reduction and calibration steps. In Section~\ref{sec:analysis}, we describe the analysis of the early-time lightcurve, while in Section~\ref{sec:models} we propose various physical models that explain it. Finally, in Section~\ref{sec:conclusions}, we discuss our findings in the context of the progenitor problem of SNe Ia, and outline our conclusions.

Throughout this paper, Modified Julian Days (MJDs) are reported as observed days while phases are reported in rest-frame, unless where noted. We adopt the AB magnitude system, unless where noted, and a Hubble constant of $H_0 = 73$ km s$^{-1}$ Mpc$^{-1}$.

\section{Discovery and Early-time Observations}
\label{sec:disc_early_phot}

SN\,2018oh was discovered by the All Sky Automated Survey for SuperNovae \citep[ASAS-SN,][]{Shappee14ApJ} in images obtained on 2018 Feb 4.41 UT (all times presented are UT) \citep[with discovery name ASASSN-18bt;][]{Brown18ATel} (Shappee et~al.\ 2018b), at $V = 15.2$~mag, with the last non-detections at 2018 Jan 27.13. The supernova is located at $\alpha=09^{\rm{h}}06^{\rm{m}}39^{\rm{s}}.592$, $\delta=+19^{\rm{o}}20\arcmin17\arcsec.47$ (J2000.0) \citep{Cornect18ATel}, $7.8\arcsec$ North and $2.0\arcsec$ East of the center of UGC~4780, a Sdm starforming galaxy, with a redshift of $z = 0.010981$ and a distance of 49.4~Mpc. The Milky Way reddening toward SN\,2018oh is $E(B-V) = 0.0368$~mag \citep{Schlafly11ApJ}. The transient was classified on 2018 Feb 5 as a relatively young ($-8$ to $-6$ days relative to peak brightness), normal SN~Ia \citep{Leadbeater18TNS, Zhang18ATel}.

UGC~4780 was included as a Campaign 16 target through `The K2 ExtraGalactic Survey (KEGS) for Transients' (PI Rest) and the  `Multi-Observatory Monitoring of K2 Supernovae' (PI Foley)  programs as part of the K2 SCE (internal Kepler ID 228682548).  After the end of Campaign 16, the data were transferred to MAST, from which we retrieved the UGC~4780 data.  We produced a provisional light curve with the ``quick look'' routine \textsc{kadenza}\footnote{\url{https://github.com/KeplerGO/kadenza}} \citep{2018ascl.soft03005B} by summing counts in a 5$\times$5 pixel aperture centered at the peak of each 30-minute image. The background was determined by estimating the median flux of the outermost pixels. Due to its unique observing strategy which requires regular thruster use to maintain pointing, K2 data suffer from a `sawtooth pattern'  and long-term sensitivity trends, partly due to temperature changes as the sun angle and the zodiacal light levels change during a Campaign. In order to correct for these effects, third-order polynomials were fit in both spatial dimensions to remove the `sawtooth.'  To account for the long-term trends, we performed a principal-component analysis that represents the common simultaneous trends seen in the light curves of all the (assumed non-varying) galaxies observed on the same chip. Through an iterative procedure, the optimal number of PCA vectors was determined to be only one. We then determined and removed the long-term trend for SN\,2018oh.  Finally, the noise was estimated by computing the root-mean-squared variation just before the explosion and then scaling this by the square root of the galaxy flux plus the SN flux in the aperture. For a more detailed discussion on the K2 reduction steps, see \citet{Shaya15AJ}.

During Campaign 16, we actively observed the K2 field with both the Pan-STARRS1 telescope \citep[PS1;][]{Chambers16arXiv,Magnier16arXiv,Waters16arXiv} and the CTIO 4-m Mayall telescope with DECam \citep{Honscheid08arXiv,Flaugher15AJ}.  The main goal was to discover and obtain multi-color light curves of transients in K2-observed galaxies.  This program was successful where we discovered 9 and 8 such transients in C16 with PS1 \citep{2018ATel11218} and DECam \citep{2018ATel11345,2018ATel11344}, respectively.  Unfortunately immediately after the explosion of SN\,2018oh, poor weather prevented observations for 7 nights.  During that gap, we did not have scheduled DECam nights either.

\begin{figure*}
\begin{center}
  \includegraphics[width=0.95\textwidth]{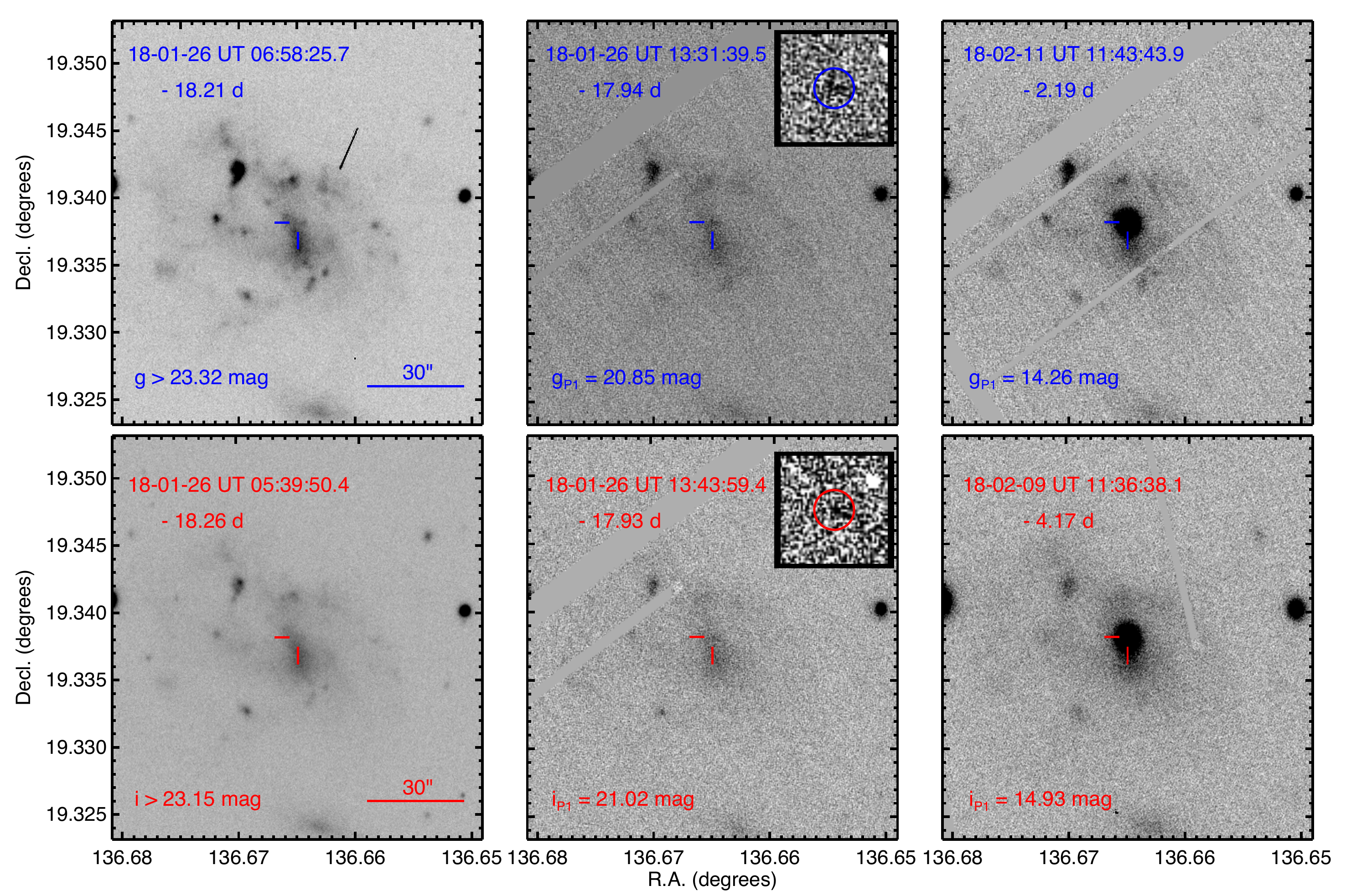}
  \caption{SN\,2018oh $g$ (top row) and $i$ (bottom row) images of pre-explosion (left), first detection (middle) and close to peak magnitude (right). The images are 107.5$\arcsec\times$107.5$\arcsec$ stamps from DECam (left) and PS1 (middle and right). For the first detections, we additionally show a zoom of 12.5$\arcsec\times$12.5$\arcsec$ of the difference image in the onset. We label the date of observation, time from $B$-band peak (in rest-frame days) and measured AB magnitude in each image stamp. The location of the SN is indicated with a tick mark (and a circle for the difference image).}
  \label{fig:SN2018oh_images}
\end{center}
\end{figure*}

All PS1 and DECam images were reduced using the \textsc{photpipe} imaging and photometry package \citep{Rest05ApJ,Rest14ApJ}, which performs standard reduction processes, including bias subtraction, cross-talk corrections, flat-fielding, astrometric calibration and image resampling. Instrumental PSF magnitudes are calculated by using \textsc{DoPhot} \citep{Schechter93PASP} on the difference images, and the final calibration is performed with PS1 standard-star fields. This photometric procedure is well-tested and has been applied in many transient studies \citep[e.g.,][]{Rest14ApJ, Kilpatrick18MNRAS}.  We present PS1 and DECam images from before and immediately after explosion, as well as images near peak brightness, in Figure~\ref{fig:SN2018oh_images}.

Spectroscopic and photometric follow-up observations of the SN were performed immediately after its discovery, and a complete presentation of the SN properties is presented in Li et~al.\ (2018).

\section{Analysis}
\label{sec:analysis}

In this Section, we present the early photometric observations of SN\,2018oh, both from ground-based facilities and K2. We then present a basic analysis of the early evolution of the SN, based on analytical models.

\subsection{Ground-based Photometry}
\label{sec:ground_based_phot}

SN\,2018oh is detected in PS1 $g$ and $i$ images, on UT 2018-01-26.56 and 26.57 (for $g$ and $i$ respectively), 8.9~days before the ASAS-SN discovery image, with AB magnitudes of $g_{\rm P1}=20.72 \pm 0.18$ and $i_{\rm P1}=20.94 \pm 0.25$, while the last non-detections were at UT 2018-01-23.38 and UT 2018-01-22.55. Moreover, from DECam $i$-band images taken one day later, SN\,2018oh was $i = 19.04 \pm 0.01$ and $18.96 \pm 0.01$~mag on 2018 Jan 27.25 and 2018 Jan 27.29, respectively, revealing a rise in the $i$ band of $\sim$1~mag in one day. A collection of ground-based images, showing pre-explosion, first detection and close-to-peak luminosity, in $g$ and $i$ bands, is presented in Figure~\ref{fig:SN2018oh_images}, and reported in Table~\ref{tab:ground_phot}.

After correcting for the Milky Way extinction using the \citet{Fitzpatrick99PASP} law with $R_{V} = 3.1$, we fit the $uBV\!griz$ photometry (Li et~al.\ 2018) with the most recent version of the \textsc{SALT2} light curve fitter (SALT2.4; \citealp{Betoule14A&A,Guy10A&A}) through the SNANA framework \citep{Kessler09PASP}.  We measure a SALT2 shape parameter of $x_{1} = 0.879 \pm 0.012$ and a color parameter of $c = -0.09 \pm 0.01$. We determine that SN\,2018oh peaked at $B_{\rm peak} = 14.185 \pm 0.010$~mag on MJD $58163.339 \pm 0.016$. 

To infer the distance, we use the distance estimator from \citet{Betoule14A&A}, and references therein:

\begin{equation}
  \mu = m_B - M_B + \alpha \times x_1 - \beta \times c + \Delta_M,
  \label{eqn:salt2}
\end{equation}

\noindent where $m_B$, $x_1$ and $c$ are given above.  We use the values of the nuisance parameters
$\alpha = 0.141$, $\beta = 3.099$ and $M_B = -19.17$ given by \citet{Betoule14A&A}. Regarding the  host galaxy mass step $\Delta_M$ \citep{Kelly10,Lampeitl10,Sullivan10}, we use SDSS $g$ and $i$ magnitudes with the relation of \citet[their Equation 8]{Taylor11} to derive the host galaxy mass of UGC~4780.  We find the mass to be 8.81 dex, comfortably on the low-mass side of the step function, and we correct with $\Delta_M=-0.06$~mag. The final distance modulus, assuming $H_0 = 73$ km s$^{-1}$ Mpc$^{-1}$, is estimated to be $\mu = 33.61 \pm 0.05$~mag, corresponding to a distance of $52.7 \pm 1.2$~Mpc.  As UGC~4780 is not in the Hubble flow and has no independent distance measurement, the distance using the SN itself is the most accurate and precise distance, and we use this distance for the remainder of the analysis.

The near-peak and post-peak photometric data of SN\,2018oh show that the SN is a normal SN~Ia, while the only spectral peculiarity is the (relatively) long-lived carbon absorption features, seen even to about 3 weeks after the maximum light and discussed in Li et~al.\ (2018). From all available data, we conclude that SN\,2018oh is a normal SN~Ia.

\subsection{\textit{Kepler} Light Curve}
\label{sec:kepler_phot}

After the reduction of the SN\,2018oh {\it Kepler}/K2 light curve as described in Section~\ref{sec:disc_early_phot}, which only provides a relative-flux light curve, we determine the true K2 flux as follows.  We use the $uBV\!griz$ photometry (Li et~al.\ 2018), which has been calibrated to the PS1 system to determine the SN\,2018oh flux as a function of time and wavelength.  We then use the ``max model" of the \texttt{SNooPy}\footnote{\url{https://users.obs.carnegiescience.edu/cburns/SNooPyDocs/html/}} package \citep{SNooPy} to determine the spectral-energy distribution (SED) of the SN as a function of time.  This model first fits for the peak flux in each photometric band by scaling template light curves \citep{Burns11AJ} to the data, with the model $K$-corrections calculated by warping the \citet{Hsiao07} SN~Ia spectral series to match the observed colors. This approach accounts for assumptions about host-reddening and the distance to the SN by modeling the multi-band photometry before determining the K2 magnitudes. The best-fit parameters were used to normalize the mangled spectral series to the observed photometry and to generate a synthetic SED. As the \citet{Burns11AJ} method mangles the spectral series to match the SED in each observer-frame passband, there is a choice of which passband's normalized SED to use as a model for the K2 band. We use the $V$ band as its effective wavelength is closest to that of the K2 band. After integrating over the K2 passband, recovering the `synthetic' K2 light curve, we solve for the absolute zeropoint, using the background-subtracted \textit{K2} flux light curve, interpolated over a range of $\pm$3~d around the time of {\it B}-band maximum light, where the supernova color evolves slowly. We estimate $\mathrm{ZP_{K2}}=25.324 \pm 0.004~\text{(statistical)}$. We find a $\pm 0.011 ~\text{(systematic)}$~mag, systematic uncertainty arising from the choice of which (observer-frame) passband normalized-SED is used to model the synthetic K2 light curve.

\begin{figure*}
\begin{center}
  \includegraphics[width=0.95\textwidth]{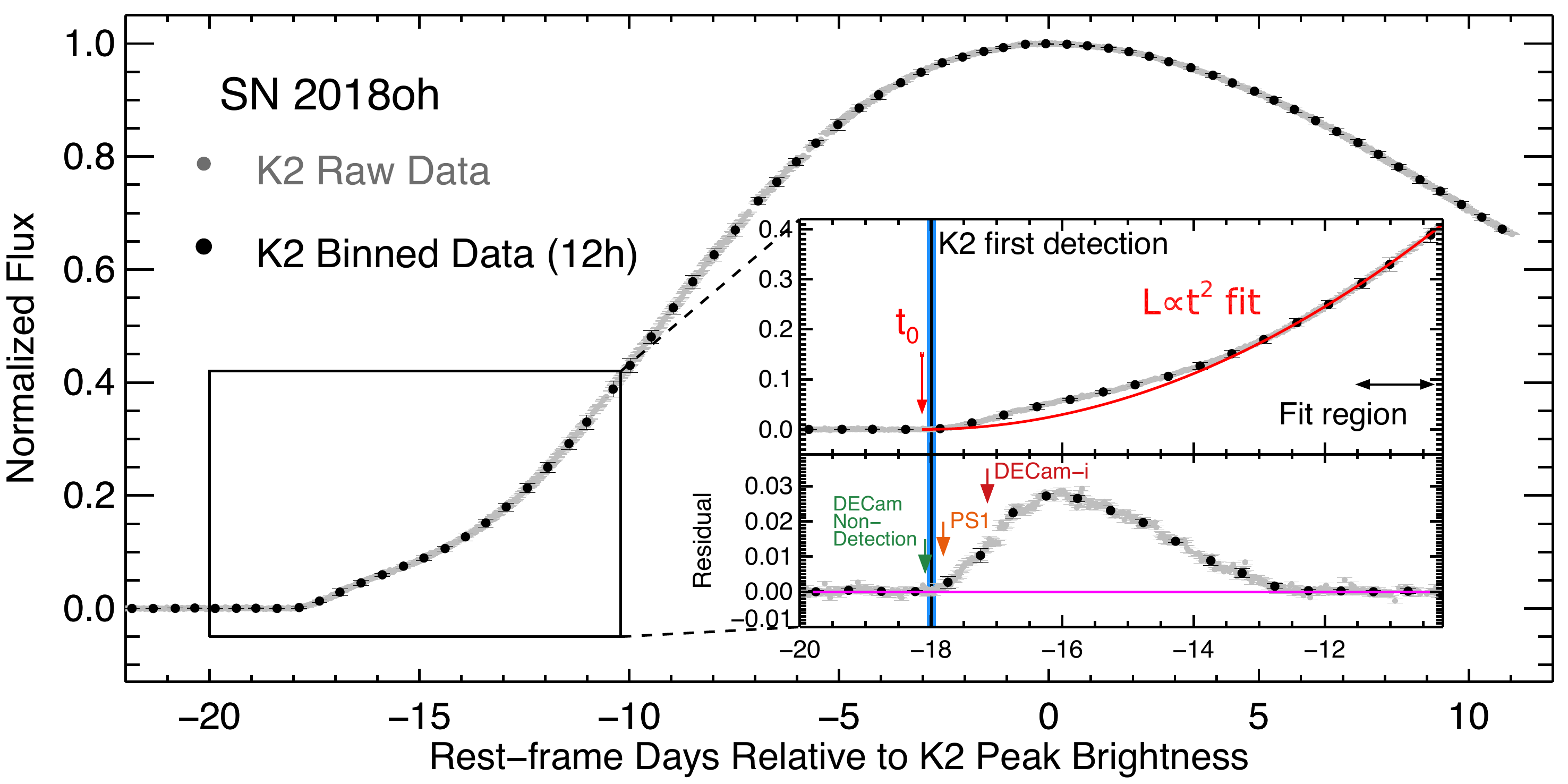}
  \caption{SN\,2018oh K2 light curve, normalized to peak flux, with respect to peak brightness. Unbinned K2 photometry and data averaged over 12 hours are shown as grey and black points, respectively. In the inset's upper panel, we show the zoomed light curve from 20 to 10~days before peak brightness. A $L \propto t^{2}$ model (red line) fit to the data in the `Fit region' is displayed.  The residual of the fit is shown in the lower panel.  The time of our last DECam non-detection, first PS1 and DECam observations are marked with green, orange and red arrows, respectively.  The black vertical line corresponds to the estimation of the onset of the K2 light curve, as described in the text, with the blue-shaded region representing the 3-$\sigma$ standard deviation.}
  \label{fig:sn2018oh_kepler_lc}
\end{center} 
\end{figure*}

We present the SN\,2018oh K2 light curve in Figure~\ref{fig:sn2018oh_kepler_lc}, normalized to the peak of the light curve, which we estimate by fitting a polynomial to the data from MJD 58160.0 to 58165.0.  We find that the peak in the K2 band occurs at $\mathrm{MJD_{max}^{\mathrm{K2}}} = 58162.58$, $\sim$0.12~days prior to $B$-band maximum, with $\mathrm{K2_{max}=14.401\pm0.001}$. A portion of the light curve is presented in Table~\ref{tab:k2_phot}, while the complete dataset is available in the electronic edition.

\subsection{Basic Analysis of the K2 Light Curve}
\label{sec:k2_phot}

Assuming that the photospheric temperature of a SN Ia does not change significantly in the first few days after explosion, the luminosity of the Rayleigh-Jeans tail of the blackbody radiation will increase with time as $L\propto t^{2}$ \citep{Arnett82ApJ,Riess99AJ}, as the size of the photosphere increases.  However, the K2 light curve of SN\,2018oh shows a prominent ``two-component rise'': an initial flux excess, from $\sim$18 to 13~days before peak brightness, which eventually subsides and the usual ``expanding fireball'' rise dominates starting about 13~days before peak brightness.

We determine the onset of the K2 light curve as follows: For a given sliding time-window, we calculate the weighted-mean of the flux and we compare it with the flux of the time-window prior to it, marking as a detection when Flux$_{i} > 3\sigma_{i-1}$. By an iterative procedure, using decreasing time-window widths, we record the detection times, and we estimate their mean and standard deviation. We calculate $t_{ \rm det}^{\rm K2}=-17.99 \pm 0.04$ days from maximum light (at MJD$_{\rm det}^{\rm K2} = 58144.39$), shown as the black vertical line in Figure~\ref{fig:sn2018oh_kepler_lc}. We note that the first PS1 detections were 0.18~days (4.32~hours) after the K2 first detection, which we estimate to be 2018 Jan 26.04. 

In order to determine the properties of the power-law rise (i.e., excluding the first-component rise), we attempt to estimate a time range by iteratively fitting, using \textsc{idl}'s MPFIT function, a $(t-t_{0})^2$ power law to the data in a window from a variable (shifting by steps of 0.02~days) start time beginning 20~days before peak brightness until the flux reaches 40\% of the peak flux, as has been done with other {\it Kepler} SN~Ia studies \citep{Olling15Natur}. Our best fit (reduced $\chi^{2} = 1.09$) is for a time window from 11.54 to 10.32~days before peak brightness, which we mark with a vertical two-headed arrow in Figure~\ref{fig:sn2018oh_kepler_lc}.  From this fit, we estimate a time of first light, $t_{0}=-18.14 \pm 0.02$~days --- $\sim$0.15~days ($\sim$3.6~hours) before our first K2 detection.  We display the residual to the fit in the bottom panel of the inset in Figure~\ref{fig:sn2018oh_kepler_lc}. We find that $\sim$2 days after $t_{0}$, the flux excess is  $\sim$3 times as luminous as the power-law rise, and represents $\sim$65\% of the total flux at that time.

As it has been shown in previous rise time studies \citep{Riess99AJ,Hayden10ApJ1,Ganeshalingam11MNRAS,GonzalezGaitan12,Firth15MNRAS}, the index of the power law can significantly vary from 2 for a particular SN. To account for this possibility, we repeat the previous procedure and, using \textsf{emcee}, a Python-based application of an affine invariant Markov chain Monte Carlo (MCMC) with an ensemble sampler \citep{Foreman-Mackey13}, we fit a $(t-t_{0})^\alpha$ power-law (thus, additionally fitting for the power-law index). Doing so, we find a similar best-fit region as before, with the new best-fit parameters $t_{0} = -17.86^{+0.24}_{-0.25}$~days before peak brightness, with $\alpha = 1.92 \pm 0.07$.

\begin{figure}
\begin{center}
  \includegraphics[width=0.45\textwidth]{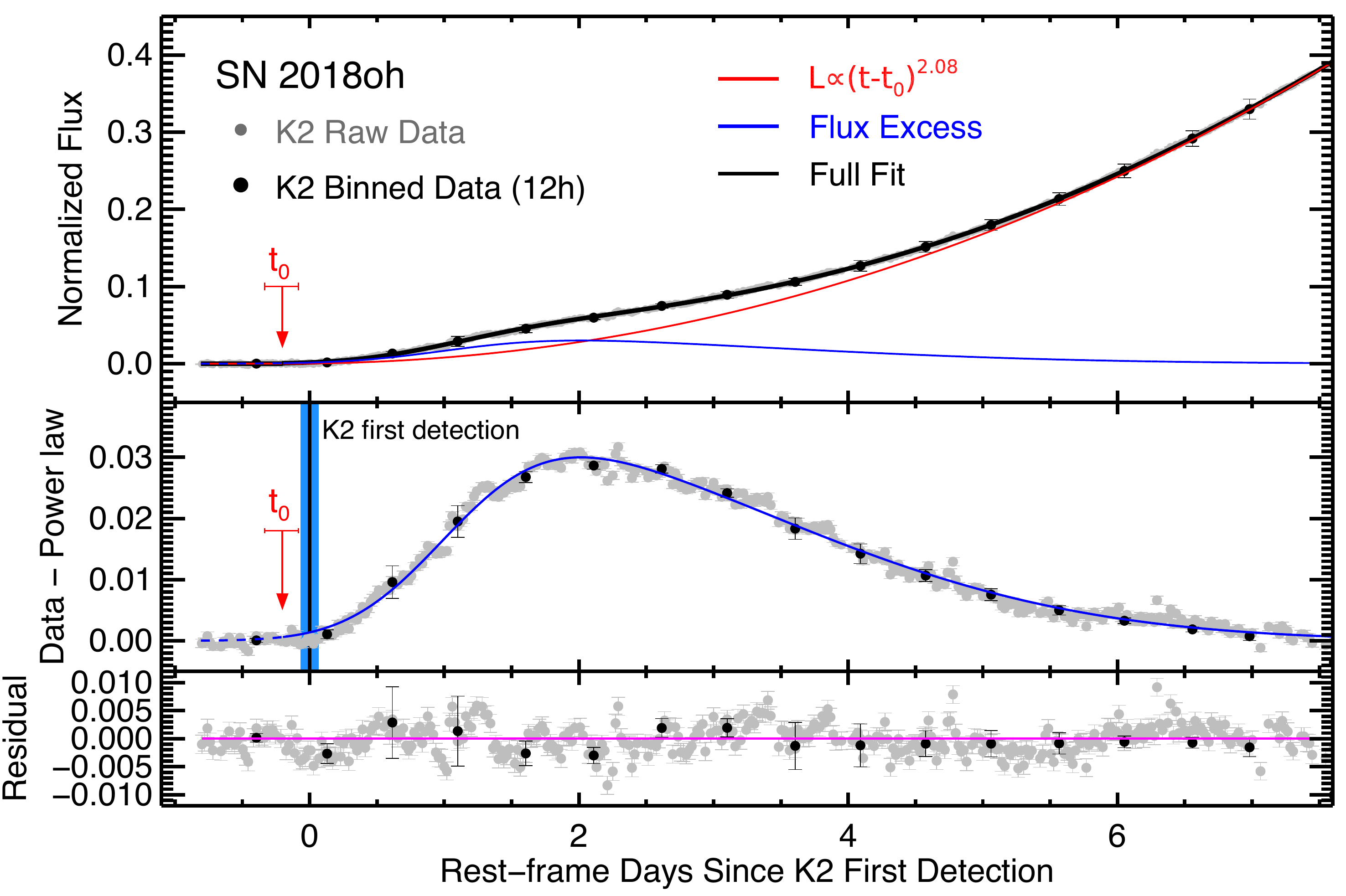}
  \caption{(Top) SN\,2018oh K2 light curve, normalized to peak flux, with respect to the first K2 detection. Our full fit is shown as a solid black line, while the decomposition of the fit is shown as a red line, for the SN power law flux, and a blue line, for the first rise component. The red downward arrow denotes the time of first light $t_{0}$, estimated from the fit. (Middle) The early flux excess, plotted as the data minus the fitted power-law model. The result of the first component fit is shown as a blue line. (Bottom) The residual of the fit (data minus full model fit).}
  \label{fig:sn2018oh_kepler_lc_fit}
\end{center} 
\end{figure}

In order to quantify the rise of the excess flux component, and motivated by its shape, we consider a simple analytical model that consists of (1) a power law $L \propto  (t-t_{0})^{\alpha}$ and (2) a skewed Gaussian to account for the early flux excess.  We fit the light curve from 20~days before peak brightness through the time when the flux reaches 40\% of the peak flux, with both a fixed power law index of 2, and with the index allowed to float. By fixing the index to $\alpha = 2$, we estimate $t_{0} = -18.00^{+0.03}_{-0.02}$.  When simultaneously fitting for the power-law index, we find $t_{0} = -18.19 \pm 0.05$ and $a = 2.08 \pm 0.02$. The later fit is shown in Figure~\ref{fig:sn2018oh_kepler_lc_fit}.  These results are generally consistent with the canonical expanding fireball model, and the initial assumption that $L \propto (t-t_{0})^{2}$ seems reasonable given the data.

From the multi-component fit, we also estimate that the early excess flux peaked with a luminosity of $(4.3 \pm 0.2)\times 10^{37}$~erg~s$^{-1}$ at $t_{\rm peak}^{c1} = -16.05 \pm 0.04$~days, approximately 2.2~days after $t_{0}$, and had a FWHM of 3.12~days.  The total emitted energy above the power-law rise is $(1.27 \pm 0.01)\times 10^{43}$~erg.

\subsection{Comparison to Other SNe}
\label{sec:phot_comp}

\begin{figure}
\begin{center}
  \includegraphics[width=0.45\textwidth]{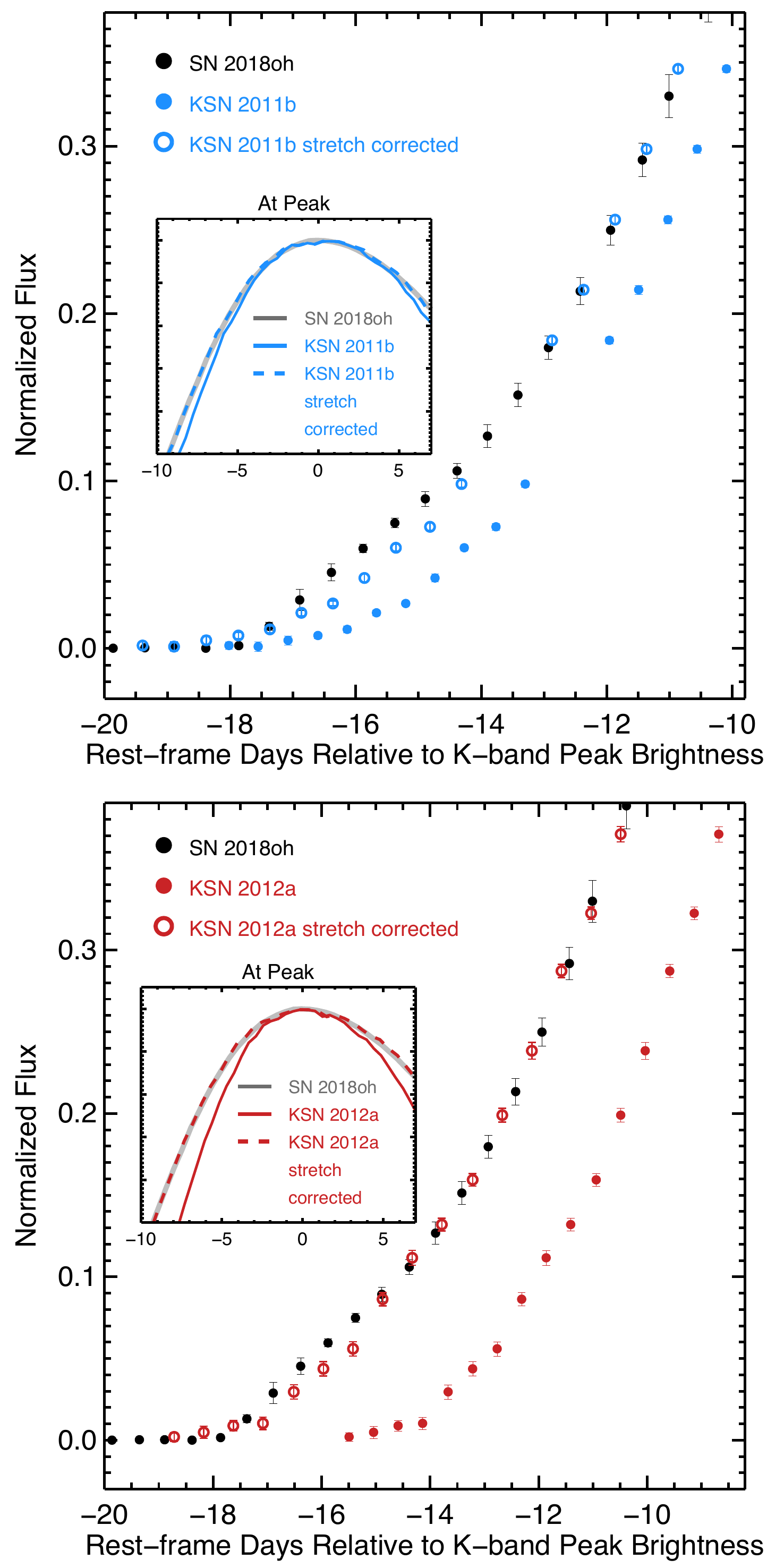}
  \caption{Comparison of the K2 SN~2018oh light curve (black circles), normalized to peak, with respect to peak brightness, with KSN 2011b (upper panel) and KSN 2012a (lower panel). The \textit{Kepler} light curves of KSN 2011b and KSN 2012a have been `stretch-corrected' to match the SN~2018oh light curve. We show the original 12-h time binned data from \citet{Olling15Natur} in full circles, and the `stretch-corrected' ones with open circles. In the insets, we show a zoom of the light curves at peak.}
\label{fig:sn2018oh_comparison_ksne}
\end{center} 
\end{figure}

Firstly, we compare the K2 light curve of SN\,2018oh with the \textit{Kepler} SNe presented in \citet{Olling15Natur}, focusing on the discovery and rise epochs (Figure~\ref{fig:sn2018oh_comparison_ksne}).  As mentioned in \citet{Olling15Natur}, KSN 2011b (blue full circles) and KSN 2012a (red full circles) occurred in red and passive galaxies at redshifts $\sim0.05$ and $\sim0.09$ (we exclude the 3rd \textit{Kepler} SN of \citealt{Olling15Natur}, KSN 2011c, due to the lower quality of data). Moreover, these SNe are fast decliners (thus, have lower absolute luminosities) while SN\,2018oh is a normal SN Ia. For this reason, we `stretch-correct' \citep{Perlmutter97ApJ} the \textit{Kepler} SN light curves to the K2 light curve of SN\,2018oh by determining the stretch factor that, when applied, best matches the light curves (see the insets in Figure~\ref{fig:sn2018oh_comparison_ksne}). The `stretch-corrected' light curves are shown as open blue (KSN 2011b) and red (KSN 2012a) circles.

As it can be seen, the applied stretch correction successfully matches the SNe at the epochs around peak brightness. However, SN\,2018oh clearly deviates for the first few days after explosion, when the flux excess is observed. We estimate that, at the time of the peak of the flux excess, $t_{\rm peak}^{c1} = -16.05$~days, SN\,2018oh is 51\% and 32\% more luminous than the stretch-corrected KSN 2011b and KSN 2012a, respectively.

\begin{figure*}
\begin{center}
  \includegraphics[width=0.95\textwidth]{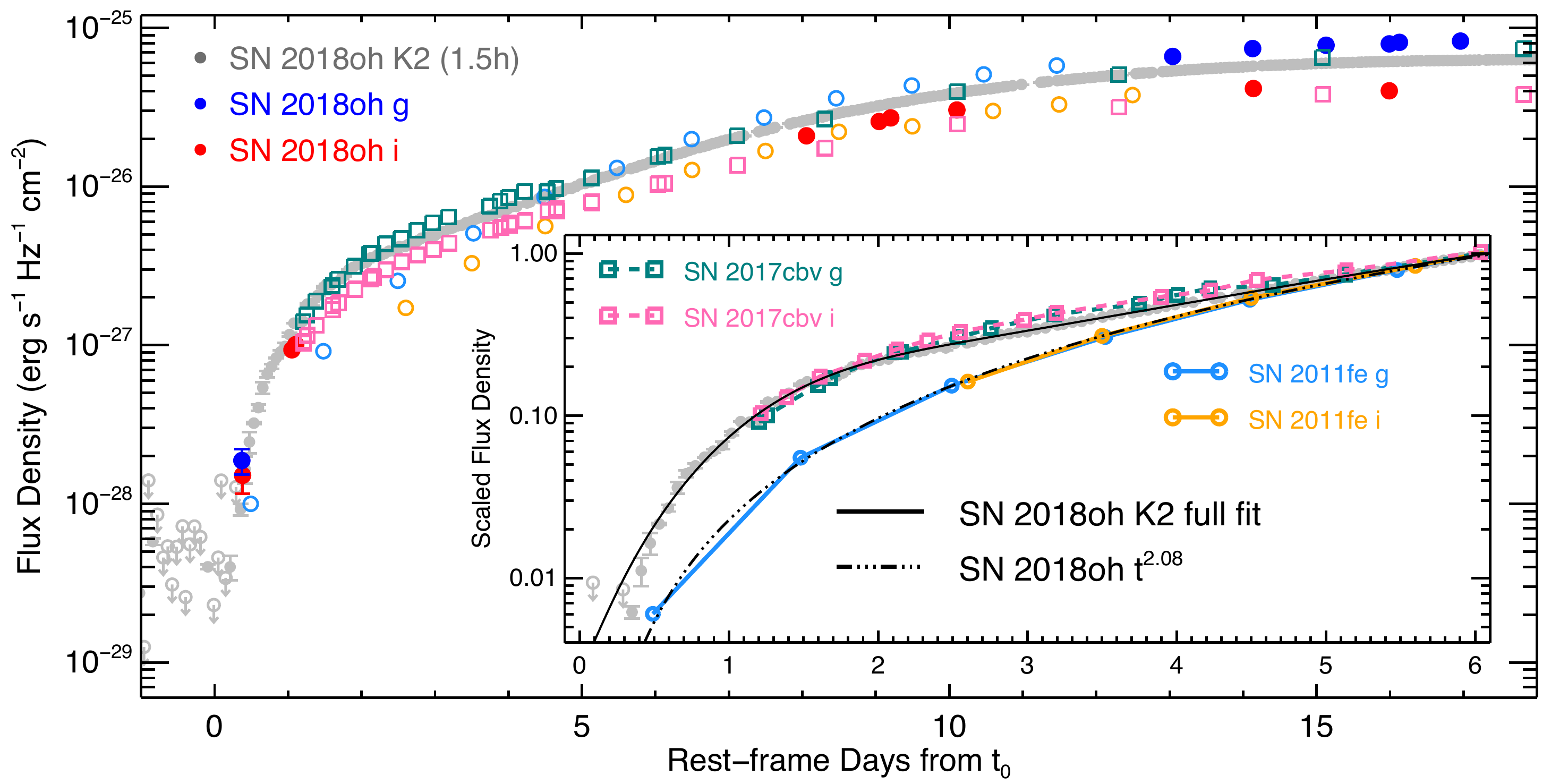}
  \caption{SN\,2018oh flux in $g_{\rm P1}$ (blue), $i$ and $i_{\rm P1}$ (red) and K2 (gray), with respect to the rest-frame time since first light, $t_{0}$, as estimated in Section~\ref{sec:k2_phot}. Non-detections in the K2 band are plotted as open gray circles. Similar light curves for SN\,2011fe ($g$ and synthetic $i_{\rm P1}$) and SN\,2017cbv ($g$ and $i$) are also shown, after being normalized to the appropriate peak flux of SN\,2018oh. (Inset) A zoomed-in region of the extremely early light curve, normalized to the K2 flux at 6~days after explosion. Note that the comparison SN sample is additionally plotted as solid (SN\,2011fe) and dashed (SN\,2017cbv) lines, with some of the photometric points removed for clarity. We overplot the fits described in Section~\ref{sec:analysis} as indicated in the legend.}
  \label{fig:sn2018oh_kepler_early_flux}
\end{center} 
\end{figure*}

Next, we compare the early SN\,2018oh light curves with those of two other SNe with very early data: the well-studied, extremely young SN\,2011fe \citep{Nugent11}, a normal Type Ia supernova that shows no flux excess at the extremely early times, and SN\,2017cbv \citep{Hosseinzadeh17ApJ}, a SN~Ia with a prominent blue early flux excess.  For this comparison, we need comparable filters.  SN\,2017cbv has extensive early-time photometry in the desired $g$ and $i$ bands \citep{Hosseinzadeh17ApJ}.  SN\,2011fe also has an early $g$-band light curve \citep{Nugent11}, but lacks an early $i$-band light curve.  In place of filtered photometry, we use the \citet{Pereira13} spectrophotometric time series, from which we synthesize an $i$-band light curve.

In Figure~\ref{fig:sn2018oh_kepler_early_flux}, we simultaneously display the early SN\,2018oh K2 (gray), $g$ (blue), and $i$ (red) light curves.  For comparison, we also show similar data for SNe\,2011fe and 2017cbv.  In the inset, we show the first 6 days after explosion, where the SN rose $>$2 orders of magnitude in flux.  We also display the full two-component fit to the SN\,2018oh light curve and just the power-law component.

While SN\,2011fe clearly lacks the flux excess of SN\,2018oh and rises close to $t^{2}$, SN\,2017cbv has a flux excess at early times and an early photometric behavior comparable to SN\,2018oh.  At later times ($t > -10$~days), all three SNe evolve similarly.  Notably, from that point on, SN\,2018oh looks identical to the ``normal'' SN\,2011fe.

\begin{figure*}
\begin{center}
  \includegraphics[width=0.95\textwidth]{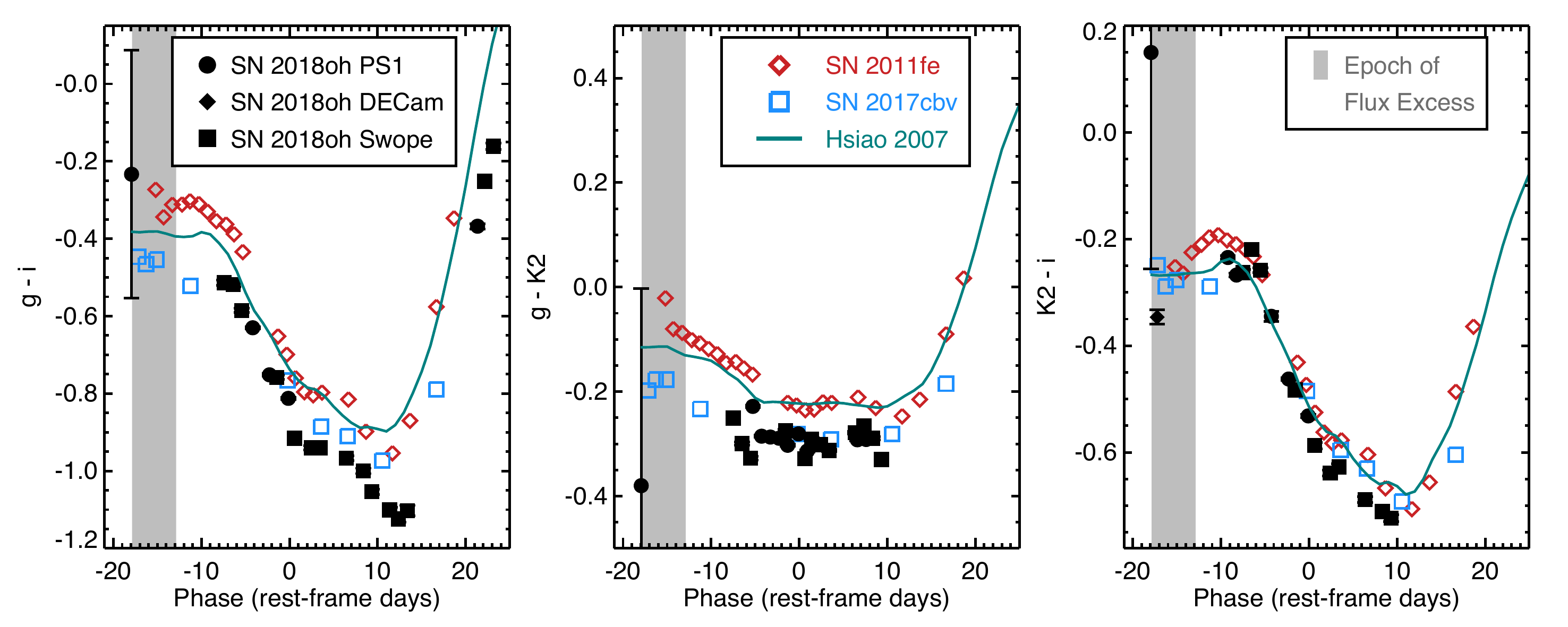}
  \caption{$g-i$ (left), $g-{\rm K2}$ (middle) and ${\rm K2}-i$ (right) color curves for SN\,2018oh. We additionally include data from the Swope telescope (Li et~al.\ 2018), to fully capture the color evolution. Similar color curves, computed as described in the text, for SN\,2011fe (red), SN\,2017cbv (blue) and the \citealt{Hsiao07} templates (green) are overplotted. The gray-shaded region corresponds to the duration of the flux excess.}
  \label{fig:sn2018oh_early_colors}
\end{center} 
\end{figure*}

Finally, we investigate the color evolution of SN\,2018oh, and in particular the $g-i$, $g-{\rm K2}$ and ${\rm K2}-i$ colors. We compare the SN\,2018oh colors to the synthetic colors of SNe\,2011fe and 2017cbv, calculated as described above (note that we also estimate the synthetic K2 magnitude). Additionally, we compute the color evolution of the \citet{Hsiao07} template spectra. The results are shown in Figure~\ref{fig:sn2018oh_early_colors}.

While SNe\,2018oh and 2017cbv generally have similar colors for the epochs examined here, generally having bluer colors than both SN\,2011fe and the \citet{Hsiao07} template, there is a distinct difference at the earliest epochs, when the prominent flux excess is observed. We note the difference in the ${\rm K2}-i$ color, at the onset of the excess-flux component where SN\,2018oh is bluer than all comparison SNe ($\sim$0.1 and 0.08 mag from SN\,2011fe and SN\,2017cbv, respectively).  Unfortunately, we only have a single $i$ observation during this phase.  Nonetheless, this single observation is critical in separating SN\,2018oh from SN\,2017cbv.

\subsection{SED of the Excess Flux}
\label{sec:sed_excess}

Finally, we investigate the SED of the excess flux observed from 18 to 13 days before peak brightness.  While we have no spectra during this phase, we have filtered photometry that can constrain the SED.  In addition to the K2 photometry, we will use the PS1 $g$ and DECam $i$ observations at $t = -17.8$ and $-17.1$~days, respectively which were obtained while the flux excess was still rising (see inset of Figure~\ref{fig:sn2018oh_kepler_lc}).

\begin{figure}
\begin{center}
  \includegraphics[width=0.45\textwidth]{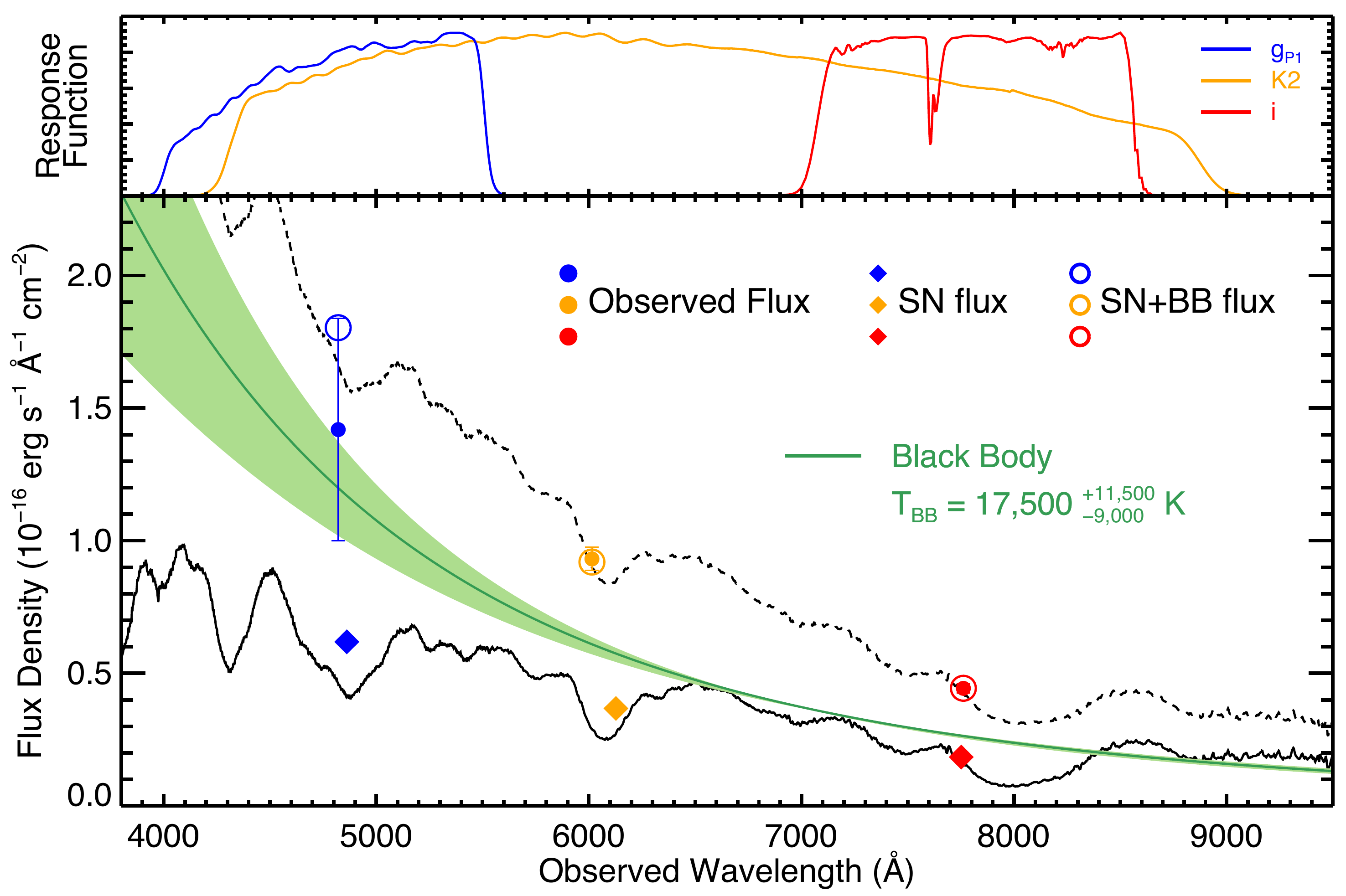}
  \caption{The +1.5~day from explosion SN\,2011fe spectrum (solid black line), redshifted and scaled to match the $t^{2.08}$ SN component of the SN\,2018oh K2 photometry at +1.09~days after $t_{0}$, obtained from the fit shown in Figure~\ref{fig:sn2018oh_kepler_lc_fit}. The $g_{\rm P1}$ (blue), K2 (orange) and $i$ (red) photometry are shown as full diamonds, at each filter's effective wavelength. The solid green line is the best-fit blackbody spectrum that reproduces the observed photometric colors at +1.09~days after $t_{0}$ (full circles), with the open circles representing the synthetic fluxes of the resulting spectrum (dashed line).  In the upper panel, we show the response functions of the $g_{\rm P1}$ (blue), K2 (orange) and DECam $i$ (red) filters.}
  \label{fig:sn2018oh_spec_match_11fe}
\end{center} 
\end{figure}

We will focus on the crucial DECam $i$ observation at $t=-17.1$ days from the K2 maximum, which coincides with the rise of the flux excess (see onset of Figure~\ref{fig:sn2018oh_kepler_lc}). While there is no spectrum of SN\,2018oh taken at that epoch, motivated by the similar peak/post-peak photometric and spectroscopic behavior with SN\,2011fe, we use the Lick/KAST spectrum, presented initially by \citet{Nugent11}, taken $\sim$1.5 days after the SN\,2011fe explosion ($-16.33$ rest-frame days from $B-$band maximum light). We attempt to spectroscopically match this spectrum (for which no flux excess is observed) with the photometric colors of SN\,2018oh at the epoch in question. As mentioned above, we unfortunately don't have $g$ observations at this epoch, therefore we assume no color evolution in $g-i$ for the first days (see left panel of Figure~\ref{fig:sn2018oh_early_colors}). We note that this assumption is somehow arbitrary: Our photometry at $-17.8$~days has large uncertainties, while the $g-i$ color is redder, compared to SN\,2017cbv. Nevertheless, after redshifting the spectrum to the redshift of SN\,2018oh, we scale it to match the SN component of the K2 flux at $-17.1$~days, as determined in Section~\ref{sec:k2_phot}, Figure~\ref{fig:sn2018oh_kepler_lc_fit}. We then perform a MCMC fit of this spectrum and a blackbody spectrum, where the resulting spectrum reproduces the observed photometry, with the results shown in Figure~\ref{fig:sn2018oh_spec_match_11fe}.

Our best fit includes a blackbody with $T=17,500^{+11,500}_{-9,000}$ K. The main source of the large uncertainty comes from the constant color evolution assumption and the corresponding large photometric uncertainty at this extremely early epoch. However, the resulting fitted temperature is high, providing an indication of a hot blackbody component, on top of the normal SN spectrum.

\section{Models} \label{sec:models}

We next consider three scenarios that may provide additional heating at early times to lead to the two-component rise seen in the SN\,2018oh light curve: the interaction between the SN and a nearby companion star, a double-detonation model with $^{56}$Ni near the surface of the star, and an additional model in which we tune that amount of surface $^{56}$Ni in an attempt to best match SN\,2018oh.

\begin{figure*}
\begin{center}
  \includegraphics[width=0.95\textwidth]{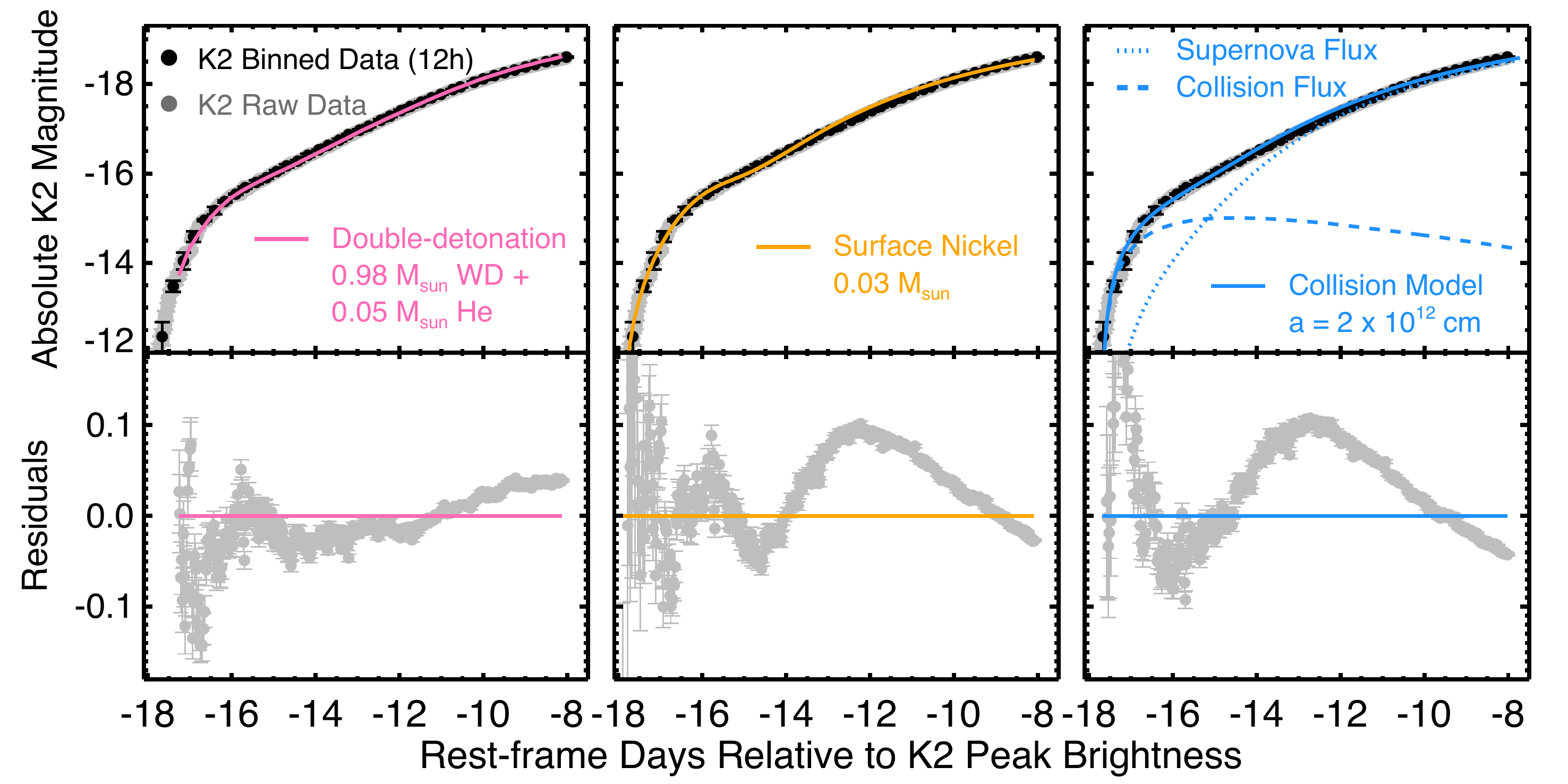}
  \caption{The absolute magnitude K2 early-time light curve of SN\,2018oh (grey and black circles). We overplot our best-fit models, as described in Section~\ref{sec:models}, of the double-detonation of a 0.98~M$_{\odot}$ WD with a 0.05~M$_{\odot}$ He layer on its surface (left panel, solid pink), of a near-Chandrasekhar mass WD explosion with a surface layer of 0.03~M$_{\odot}$ Nickel (middle panel, solid orange) and of a collision model \citep{Kasen10ApJ} with a non-degenerate companion at a distance of $a=2\times10^{12}$~cm (right panel, solid blue). We show the decomposition of the collision model to the contribution of the SN flux (dotted blue) and the interaction of the SN ejecta with the companion (dashed blue). The residuals of the model fits are shown in the lower panels.}
  \label{fig:sn2018oh_kepler_models}
\end{center} 
\end{figure*}

\subsection{Interaction with a companion star}
\label{sec:interaction_kasen}

One potential explanation for the early-time excess is shock-interaction between the supernova ejecta and a non-degenerate binary companion \citep{Kasen10ApJ}. The collision is characterized by prompt X-ray emission, followed by an optical/UV excess lasting about one week after explosion. Although the excess peaks in the UV, a measurable signature is observable in the \textit{Kepler} bandpass if the system is configured in a favorable viewing angle \citep{Olling15Natur}.

To test this scenario, we use a numerical model for the early light curve following the methods outlined in \citet{Piro16ApJ} (using the Chandrasekhar progenitor models from \citealp{Martinez16}) that roughly matches the rise of SN\,2018oh, and then combine this with the analytic interaction model of \citet{Kasen10ApJ}. The interaction emission is mainly controlled through two parameters, the orbital separation $a$ and the characteristic ejecta velocity $v$. Since $a$ can vary by many orders of magnitude and $v$ is relatively well constrained to be $v\approx 10^9\,{\rm cm\,s^{-1}}$, this makes the interaction a powerful diagnostic for measuring $a$. In addition, there are viewing angle effects, but this is somewhat degenerate with the other parameters. Thus we focus on the the case when one is observing directly into the shocked region (when the companion is roughly between the explosion and the observer) and take the measured $a$ as a lower limit to the orbital separation.

From this procedure, we find that a collision with a companion at $a=2\times10^{12}\,{\rm cm}$ provides a reasonable match to the early rise. We plot this as a solid blue line in the right panel of Figure \ref{fig:sn2018oh_kepler_models}, and also show the constituent parts of the interaction (dashed blue line) and the SN itself (dotted blue line). An important assumption of this model is that the companion is overfilling its Roche lobe and therefore we can approximate its radius as \citep{Eggleton83}
\begin{equation}
	R = \frac{0.49q^{2/3}}{0.6q^{2/3}+\ln(1+q^{1/3})}a,
\end{equation}
where $q$ is the ratio of the companion and WD's mass. For a range of companion masses from $M\approx1-6\,$M$_\odot$, this results in $R\approx10-15\,$R$_\odot$, respectively. This is generally too large for a main sequence star, and thus we conclude that the companion must be a subgiant if interaction is the correct explanation for the early excess.

\subsection{Double-detonation sub-Chandrasekhar explosion}
\label{sec:abis_doubledetonations}

Another possible mechanism for creating an early-time flux excess is the double-detonation scenario for exploding a sub-Chandrasekhar mass C/O WD with an accreted shell of helium on its surface.  In this scenario, the helium shell detonates, producing on the surface some abundance of radioactive elements such as $^{56}$Ni and $^{48}$Cr, and sending a shockwave into the WD that then ignites the C/O core \citep{1994ApJ...423..371W}.  The result produces observables generally consistent with a SN~Ia, however, the amount of Fe-group elements synthesized during the He-shell detonation must be small to resemble SNe~Ia near peak brightness.  The photons produced by the radioactive decay of material on the surface quickly diffuse out of the ejecta, creating a flux excess relative to a typical SN~Ia in the first few days after explosion \citep{Noebauer17MNRAS}. 

We test this scenario as a candidate for SN\,2018oh by exploring a hydrodynamic and radiative transfer numerical survey of double-detonations of sub-Chandrasekhar mass white dwarfs, the results of which are presented in \citet{Polin18arXiv}.  The parameter space of the survey spans from $0.7-1.2\,\mathrm{M_{\odot}}$ WDs with helium shells from 0.01 to 0.08 $\mathrm{M_{\odot}}$ and a range of mixing mass from 0.05 to 0.3 $\mathrm{M_{\odot}}$. The best-fitting model, based on a reduced $\chi^{2}$ measurement, is a 0.98~M$_\odot$ WD with 0.05~M$_\odot$ of Helium on its surface, with the ejecta smoothed over a mixing length of 0.25 M$_\odot$. This model produces a total of 0.448 $\mathrm{M_{\odot}}$ of $^{56}$Ni, 3.65$\times10^{-3}$ $\mathrm{M_{\odot}}$ of $^{48}$Cr and 1.8$\times10^{-2}$ $\mathrm{M_{\odot}}$ of $^{52}$Fe. From these elements, the amount of each that is synthesized in the helium shell (i. e., in the outer layers of ejecta) is 1.22$\times10^{-2}$, 3.19$\times10^{-2}$ and 6.11$\times10^{-2}$ $\mathrm{M_{\odot}}$, corresponding to 2.7, 87.4 and 33.9\%, respectively. The K2 synthetic light curve is shown in the left panel of Figure~\ref{fig:sn2018oh_kepler_models}. The approximate magnitudes of both the early-time excess and peak are reproduced, as is the duration of the excess and rise time to peak brightness.

\subsection{A general off-center nickel distribution}
\label{sec:offcenter_nickel}

The previous model is specifically applicable to the DD scenario, but it is possible in principle that other scenarios may mix $^{56}$Ni to the outermost layers. To explore this possibility more generally, we consider models in which we take a normal SN Ia explosion and place by hand some amount of $^{56}$Ni near the surface. As with the supernova model for the interaction scenario, we use the methods outlined in \citet{Piro16ApJ} with the progenitors generated in the work of \citet{Martinez16}. Using this we place the $^{56}$Ni in two distinct regions, a centrally concentrated region that provides the main rise and a shallow region above a mass coordinate of $1.3\,$M$_\odot$. The shallow abundance is varied to find the best fit with the K2 photometry, including smoothing with a $0.05\,$M$_\odot$ boxcar which prevents numerical issues from sharp compositional gradients.

Our best-fit model under this scenario has $0.03\,$M$_{\odot}$ of $^{56}$Ni near the surface of the WD as shown by the orange solid line in the middle panel of Figure~\ref{fig:sn2018oh_kepler_models}.  The model reproduces the general evolution of the light curve, encapsulating the initial flux excess.  We therefore provisionally consider this a viable model. Whether or not such a model can reproduce the full photometric and spectroscopic evolution of SN\,2018oh is less clear. Iron-peak elements at shallow depths can provide extensive line-blanketing that alters the colors and spectra of the SN at peak luminosity, potentially making it difficult for SN\,2018oh to be a spectroscopically normal SN Ia. Below we consider in further detail whether such a model can even reproduce the early color evolution of SN\,2018oh.

\subsection{Detailed Model Comparisons}
\label{sec:model_comparisons}

Having found both SD and DD models that can reproduce the K2 light curve, we must examine additional data that differentiate these scenarios.  The earliest detections by PS1 and DECam are particularly powerful for this purpose.

In addition to detecting the flux excess in the K2 band, we also detect an excess in the $i$ band (see Figure~\ref{fig:sn2018oh_kepler_early_flux}). Examining the ${\rm K2} - i$ color during the flux excess,  we find that SN\,2018oh is bluer than SN\,2011fe by almost 0.2~mag.  Moreover, SN\,2018oh is also similarly bluer than SN\,2017cbv at that epoch. This means that one day after explosion, SN\,2018oh is not only distinct from the normal SN\,2011fe but also from SN\,2017cbv, which also had an early-time flux excess \citep{Hosseinzadeh17ApJ}. After the onset of the canonical SN rise (rightwards of the grey-shaded region in the panels of Figure~\ref{fig:sn2018oh_early_colors}), the three SNe evolve in a similar manner (apart from the usual color dispersion seen in Type Ia supernovae).

All models examined above are able to reproduce the flux excess at early times of SN\,2018oh, but with two main different physical origins.  These models predict very different SEDs and in particular different colors.  Specifically the companion-interaction model is expected to be bluer than the surface-Ni model.

\begin{figure*}
\begin{center}
  \includegraphics[width=0.95\textwidth]{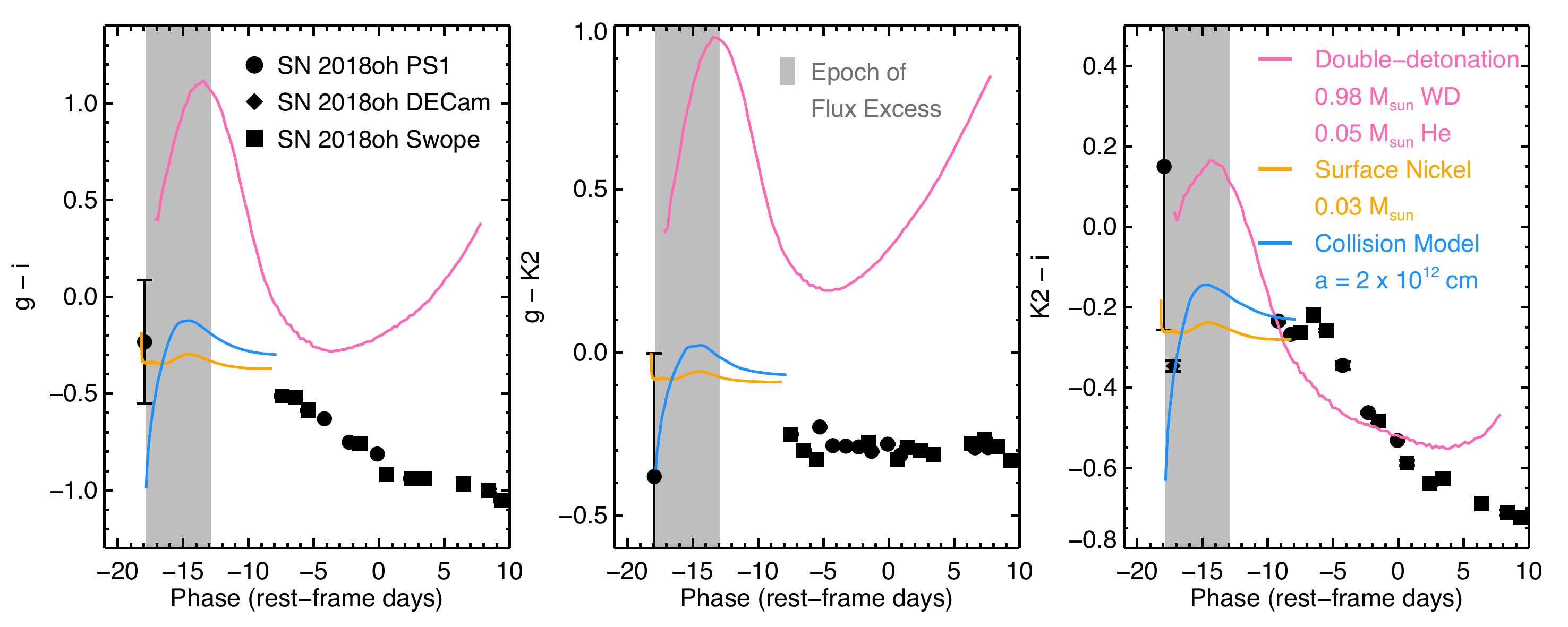}
  \caption{Same as Figure~\ref{fig:sn2018oh_early_colors} (note the different ranges in the axes), but comparing with the predicted colors of the double-detonation model (pink), the Nickel-surface model (orange) and the collision model (blue).}
  \label{fig:sn2018oh_model_colors}
\end{center} 
\end{figure*}

We display the expected $g - i$, $g - {\rm K2}$ and ${\rm K2} - i$  colors for each model in Figure~\ref{fig:sn2018oh_model_colors}.  The double-detonation model is particularly red, and it is unable to explain the blue colors of SN\,2018oh. Moreover, sub-Chandrasekhar double-detonations also leave spectral signatures such as a significant \ion{Ti}{2} absorption features in the peak spectra, that are not seen in SN\,2018oh (Li et~al.\ 2018). The generic off-center $^{56}$Ni model also has a color that is redder than SN\,2018oh by $\sim$0.1 mag.  On the other hand, the companion-interaction model with a companion at $\approx2\times10^{12}$~cm generally matches both the early rise and the color evolution of SN\,2018oh. 

Based on the color evolution of SN\,2018oh, we slightly favor the companion-interaction model over other models that can also reproduce the early flux excess.  A direct prediction of this model is the presence of hydrogen and/or helium-rich material stripped from the companion star at the nebular phases \citep{Pan12ApJ,Liu13ApJ}. To this end, detections of H or He features at late-time spectra of SN\,2018oh is crucial to confirm this model.

We note that, while SN\,2018oh has an exquisite K2 light curve, we lack the detailed color information to conclusively decide between models, particularly at bluer wavelengths. Additionally there are no spectra of SN\,2018oh during the flux excess, which would have provided key information for understanding how spectral features affect the color evolution. We do not see an abrupt flux excess on the first rise component, but rather a linear rise similar to other SNe Ia with sufficiently early, high-cadence photometry that also have two-component rising light curves \citep[e.g., SN\,2012fr, which arguably shows no signs of interaction;][]{Contreras18ApJ...859...24C} argues against the interaction model. A linear rise may result from a particular interaction model (i.e., specific viewing angle, radius, and/or separation), however an exploration of how different parameter combinations affect the detailed rise behavior is beyond the scope of this paper.

However, another interpretation of the flux excess is possible. Given the growing sample of SNe~Ia that show two-component early light curves, with different slopes and durations, the distinct early light curve evolution of SN\,2018oh, compared to the total SN~Ia population, could not be due to some external heating source, but rather a reflection of  varying SNe Ia properties, such as the density profile of the ejecta, different composition/metallicity of the progenitor star, asymmetries during the explosion etch \citep[see][for a relevant discussion]{Stritzinger18arXiv180707576S}. Modern transient surveys, such as ASAS-SN and ZTF, and future powerful surveys, such as LSST, will discover very young SNe~Ia, and with rapid follow-up, the early-light curve SN sample will increase, allowing us to investigate this possibility.

\section{Discussion and Conclusions} 
\label{sec:conclusions}

In this paper, we analyze the early photometric observations of SN\,2018oh, a normal SN~Ia, that occurred within the \textit{Kepler} Campaign 16 field.  The SN was observed with unprecedented cadence by the K2 SCE with complementary early Pan-STARRS1 and DECam imaging.  The combination of an extremely early detection and unprecedented continuous coverage with \textit{Kepler} make SN\,2018oh a spectacular reference object for early SN~Ia studies, providing invaluable insights on the explosion physics and the progenitor system. 

In the SN\,2018oh K2 and $i$-band light curves, we detect a distinct flux component in the first few days after explosion relative to other well-observed SNe~Ia (e.g., SN\,2011fe) and the $t^{2}$ luminosity rise seen later in the evolution of SN\,2018oh.  This flux excess lasts approximately 5 days, after which SN\,2018oh appears to evolve in a fashion consistent with typical SNe~Ia.

Our work provides new insights on the early time evolution of SNe~Ia, for which we find the following:

\begin{enumerate}  
\item The early K2 light curve shows a distinct two-component rise evolution.  Initially, the SN rises quickly, with a steep linear gradient, in flux.  This flux subsides after about 5 days, when a $L \propto t^{2}$ rise dominates the SN evolution.
\item Using the $t^{2}$ component of the K2 light curve, we constrain $t_{0}$ to be $-18.19 \pm 0.09$~days before K2 peak brightness.  This time is consistent with the onset of the flux excess, indicating that both components began at the same time.
\item Assuming that the $t^{2}$ component exists while the other component is bright, we find that the early flux excess peaks 2.14~days after $t_{0}$, has a FWHM of 3.12~days, a blackbody temperature of 17,500~K, a peak luminosity of 4.3$\times10^{37}$ erg/s, and a total emitted energy of 1.3$\times10^{43}$ erg.
\item We observed SN\,2018oh with Pan-STARRS1 and DECam only 4.1 and 20.6~hours after the first K2 detection, respectively, providing some of the earliest colors of a SN~Ia ever. The flux excess is confirmed in our ground-based $i$ light curve.
\item The SN\,2018oh early photometric evolution is relatively similar to SN\,2017cbv, another SN~Ia with a prominent two-component rising light curve.  However, SN\,2018oh shows bluer ${\rm K2} - i$ colors than SN\,2017cbv.  This is especially true during the epoch of the initial flux excess.  Around peak brightness, SN\,2018oh is similar to both SNe\,2011fe and 2017cbv.

\item The early flux excess can potentially be explained by additional heating at the epoch in question. We investigate three possible sources:

\begin{enumerate}
\item The interaction with a non-degenerate companion at $a=2\times10^{12}\,{\rm cm}$, with a $M\approx 1-6\,$M$_\odot$ Roche-lobe-filling star.
\item The presence of a $0.05\,\mathrm{M_{\odot}}$ Helium shell on the surface of $0.98\,\mathrm{M_{\odot}}$ C/O WD, and a subsequent sub-Chandrasekhar mass explosion.
\item An off-center $^{56}$Ni distribution of 0.03 $\mathrm{M_{\odot}}$.
\end{enumerate}

All of these models can, generally, reproduce the early shape of the K2 light curve. We slightly favor the interaction scenario, due to the blue colors at the epoch of the flux excess. However, another interpretation of the flux excess considers an intrinsic variation of early time behavior, due to varying SNe Ia properties, with no external heating source required.

\end{enumerate}

While a SD origin for (at least a sizable fraction of) SNe~Ia has been previously proposed \citep{Sternberg11Sci,Foley12ApJ2,Maguire13MNRAS}, its validity has been questioned.  Most SNe that have some observational evidence for the presence of a non-degenerate companion are either peculiar \citep[e.g., SNe~Ia-CSM;][]{Dilday12Sci,Silverman13ApJS,Fox15MNRAS.447..772F} or have contradictory observations \citep{Shappee18ApJ,Sand18arXiv}.  The general progenitor picture that has emerged over the last decade is that most SNe~Ia have a DD origin \citep{Maoz14ARA&A}.

Excluding the early-time flux excess, SN\,2018oh shows no signs of photometric and spectroscopic peculiarities.  Therefore, SN\,2018oh represents a normal SN~Ia with a potential SD origin, challenging the idea that all normal SNe~Ia have DD progenitors.  Additional SNe~Ia observed at high cadence during the first few days after explosion are needed to determine the fraction of SNe~Ia with SD progenitors.  At the same time, these observations will grow the early light curve SN Ia sample, and investigate correlations of the light curve evolution with various SNe Ia properties.

The K2 SCE has finished and the data are currently collected and analysed. With some luck, we will soon have additional K2-observed SNe~Ia with data similar in quality to that of SN\,2018oh.  

We will continue to monitor SN\,2018oh.  Late-time observations, after the SN has become optically thin, will be a direct test of our proposed models.

\begin{acknowledgments} 

\bigskip

We thank the anonymous referee for helpful comments that improved the clarity and presentation of this paper.

This paper includes data collected by the K2 mission. Funding for the K2 mission is provided by the NASA Science Mission directorate.

KEGS is supported in part by NASA K2 cycle 4 and 5 grants NNX17AI64G and 80NSSC18K0302, respectively. AR and his groups is supported in part by HST grants GO-12577 and HST AR-12851.

Pan-STARRS (PS1) is supported in part by the National Aeronautics and Space Administration under Grants NNX12AT65G and NNX14AM74G.  The Pan-STARRS1 Surveys (PS1) and the PS1 public science archive have been made possible through contributions by the Institute for Astronomy, the University of Hawaii, the Pan-STARRS Project Office, the Max-Planck Society and its participating institutes, the Max Planck Institute for Astronomy, Heidelberg and the Max Planck Institute for Extraterrestrial Physics, Garching, The Johns Hopkins University, Durham University, the University of Edinburgh, the Queen's University Belfast, the Harvard-Smithsonian Center for Astrophysics, the Las Cumbres Observatory Global Telescope Network Incorporated, the National Central University of Taiwan, the Space Telescope Science Institute, the National Aeronautics and Space Administration under Grant No.\ NNX08AR22G issued through the Planetary Science Division of the NASA Science Mission Directorate, the National Science Foundation Grant No.\ AST--1238877, the University of Maryland, Eotvos Lorand University (ELTE), the Los Alamos National Laboratory, and the Gordon and Betty Moore Foundation.

This project used data obtained with the Dark Energy Camera (DECam), which was constructed by the Dark Energy Survey (DES) collaboration. Funding for the DES Projects has been provided by the DOE and NSF (USA), MISE (Spain), STFC (UK), HEFCE (UK), NCSA (UIUC), KICP (U. Chicago), CCAPP (Ohio State), MIFPA (Texas A\&M), CNPQ, FAPERJ, FINEP (Brazil), MINECO (Spain), DFG (Germany) and the collaborating institutions in the Dark Energy Survey, which are Argonne Lab, UC Santa Cruz, University of Cambridge, CIEMAT-Madrid, University of Chicago, University College London, DES-Brazil Consortium, University of Edinburgh, ETH Z\"urich, Fermilab, University of Illinois, ICE (IEEC-CSIC), IFAE Barcelona, Lawrence Berkeley Lab, LMU M\"unchen and the associated Excellence Cluster Universe, University of Michigan, NOAO, University of Nottingham, Ohio State University, OzDES Membership Consortium, University of Pennsylvania, University of Portsmouth, SLAC National Lab, Stanford University, University of Sussex, and Texas A\&M University.

Based on observations at Cerro Tololo Inter-American Observatory, National Optical Astronomy Observatory (NOAO 2017B-0279; PI: A Rest, NOAO 2017B-0285; PI: A Rest), which is operated by the Association of Universities for Research in Astronomy (AURA) under a cooperative agreement with the National Science Foundation.

The UCSC group is supported in part by NASA grants NNG17PX03C and 80NSSC18K0303, NSF
grants AST-1518052 and AST-1815935, the Gordon \& Betty Moore Foundation, the
Heising-Simons Foundation, and by fellowships from the Alfred P.\
Sloan Foundation and the David and Lucile Packard Foundation to R.J.F.

We thank Chris Burns for providing crucial updates to \texttt{SNooPy}, relevant to the analysis of this work. 

SJS acknowledges funding from STFC Grants ST/P000312/1 and ST/N002520/1.

This work makes use of observations from Las Cumbres Observatory. DAH, CM, and GH are supported by the US National Science Foundation grant 1313484. Support for IA was provided by NASA through the Einstein Fellowship Program, grant PF6-170148.

JV and his group at Konkoly Observatory is supported by the project ``Transient Astrophysical Objects" GINOP 2.3.2-15-2016-00033 of the National Research, Development and Innovation Office (NKFIH), Hungary, funded by the European Union. 

This project has been supported by the Lend\"ulet Program of the Hungarian Academy of Sciences, projects No. LP2018-7/2018 and LP2012-31, and the NKFIH grant K-115709.

ZsB acknowledges the support provided from the National Research, Development and Innovation Fund of Hungary, financed under the PD 17 funding scheme, project no. PD123910.

Research by DJS  is supported by NSF grants AST-1821967, 1821987, 1813466 and 1813708. NS and JEA received support from NSF grant AST-1515559.

We acknowledge the support of the staff of the Lijiang 2.4m and Xinglong 2.16m telescope. Funding for the LJT has been provided by Chinese Academy of Sciences and the People's Government of Yunnan Province. The LJT is jointly operated and administrated by Yunnan Observatories and Center for Astronomical Mega-Science, CAS. This work is supported by the National Natural Science Foundation of China (NSFC grants 11178003, 11325313, and 11633002), and the National Program on Key Research and Development Project (grant no. 2016YFA0400803). JJZ is supported by the National Science Foundation of China (NSFC, grants 11403096, 11773067), the Youth Innovation Promotion Association of the CAS, the Western Light Youth Project, and the Key Research Program of the CAS (Grant NO. KJZD-EW-M06). TMZ is supported by the NSFC (grants 11203034). This work was also partially Supported by the Open Project Program of the Key Laboratory of Optical Astronomy, National Astronomical Observatories, Chinese Academy of Sciences. 

EB and JD acknowledge partial support from NASA grant NNX16AB5G.

CPG acknowledges support from EU/FP7-ERC grant No. [615929].

Parts of this research were supported by the Australian Research Council Centre of Excellence for All Sky Astrophysics in 3 Dimensions (ASTRO 3D), through project number CE170100013.

Support for this work was provided by NASA through Hubble Fellowship grant \#HST-HF2-51357.001-A, awarded by the Space Telescope Science Institute, which is operated by the Association of Universities for Research in Astronomy, Incorporated, under NASA contract NAS5-26555, as well as NASA {\em K2} Cycle 4 Grant NNX17AE92G.

\end{acknowledgments}

\appendix

\section{Photometry Tables}

\begin{deluxetable*}{lcDDCC}[t!]
\tablecaption{SN~2018oh Ground-based Photometry \label{tab:ground_phot}}
\tablecolumns{6}
\tablenum{1}
\tablewidth{0pt}
\tablehead{
\colhead{UT Date} &
\colhead{MJD} &
\twocolhead{Phase\tablenotemark{a}}  & \twocolhead{Time from detection\tablenotemark{b}}  & \colhead{Filter} & \colhead{Value} \\
\colhead{(YYYY-mm-dd)} & \colhead{(Days)} & \twocolhead{(Rest-frame days)} & \twocolhead{(Rest-frame days)} & \colhead{} & \colhead{(Mag)}
}
\decimals
\startdata
2018-01-26.24 & 58144.24 & \phn \phn \phn -18.26 & \phn \phn \phn \phn \phn \phn -0.15 & i & 23.155$^{c}$ \\
2018-01-26.29 & 58144.29 & -18.21 & -0.10 & g & 23.322$^{c}$ \\
2018-01-26.56 & 58144.56 & -17.94 & 0.17 & g_{\rm P1} & 20.852 \pm 0.224 \\
2018-01-26.57 & 58144.57 & -17.93 & 0.18 & i_{\rm P1} & 21.022 \pm 0.268 \\
2018-01-27.25 & 58145.25 & -17.26 & 0.85 & i & 19.039 \pm 0.013 \\
2018-01-27.30 & 58145.30 & -17.21 & 0.90 & i & 18.957 \pm 0.014 \\
2018-02-03.33 & 58152.33 & -10.26 & 7.85 & i_{\rm P1} & 15.670 \pm 0.004 \\
2018-02-04.33 & 58153.33 & -9.27 & 8.84 & i_{\rm P1} & 15.445 \pm 0.004 \\
2018-02-04.49 & 58153.49 & -9.11 & 9.00 & i_{\rm P1} & 15.388 \pm 0.003 \\
2018-02-05.40 & 58154.40 & -8.21 & 9.90 & i_{\rm P1} & 15.264 \pm 0.010 \\
2018-02-08.37 & 58157.37 & -5.27 & 12.84 & g_{\rm P1} & 14.483 \pm 0.002 \\
2018-02-09.47 & 58158.47 & -4.18 & 13.93 & g_{\rm P1} & 14.359 \pm 0.002 \\
2018-02-09.48 & 58158.48 & -4.17 & 13.94 & i_{\rm P1} & 14.926 \pm 0.003 \\
2018-02-10.48 & 58159.48 & -3.18 & 14.93 & g_{\rm P1} & 14.307 \pm 0.002 \\
2018-02-11.35 & 58160.35 & -2.32 & 15.79 & g_{\rm P1} & 14.285 \pm 0.002 \\
2018-02-11.35 & 58160.35 & -2.32 & 15.79 & i_{\rm P1} & 14.961 \pm 0.003 \\
2018-02-11.49 & 58160.49 & -2.19 & 15.92 & g_{\rm P1} & 14.261 \pm 0.002 \\
2018-02-12.33 & 58161.33 & -1.36 & 16.76 & g_{\rm P1} & 14.241 \pm 0.002 \\
2018-02-13.56 & 58162.56 & -0.14 & 17.97 & g_{\rm P1} & 14.253 \pm 0.002 \\
2018-02-13.57 & 58162.57 & -0.13 & 17.98 & i_{\rm P1} & 15.002 \pm 0.003 \\
2018-02-14.53 & 58163.53 & 0.82 & 18.93 & g_{\rm P1} & 14.225 \pm 0.002 \\
2018-02-20.34 & 58169.34 & 6.57 & 24.68 & g_{\rm P1} & 14.418 \pm 0.002 \\
2018-02-21.49 & 58170.49 & 7.71 & 25.82 & g_{\rm P1} & 14.472 \pm 0.002 \\
2018-03-07.28 & 58184.28 & 21.35 & 39.46 & g_{\rm P1} & 15.465 \pm 0.005 \\
2018-03-07.29 & 58184.29 & 21.36 & 39.47 & i_{\rm P1} & 15.769 \pm 0.006 \\
2018-03-08.25 & 58185.25 & 22.30 & 40.42 & g_{\rm P1} & 15.543 \pm 0.004 \\
2018-03-08.41 & 58185.41 & 22.46 & 40.57 & g_{\rm P1} & 15.514 \pm 0.003 \\
2018-03-18.29 & 58195.29 & 32.24 & 50.35 & i_{\rm P1} & 15.635 \pm 0.006 \\
2018-03-18.29 & 58195.29 & 32.24 & 50.35 & g_{\rm P1} & 16.306 \pm 0.009 \\
\enddata
\tablenotetext{a}{Relative to MJD$_{B}^{\rm peak} = 58162.7$}
\tablenotetext{b}{Relative to MJD$_{\rm detection}^{\rm K2} = 58144.39$}
\tablenotetext{c}{3$\sigma$ upper limit}
\end{deluxetable*}

\begin{deluxetable*}{lcDDCC}[t!]
\tablecaption{SN~2018oh K2 Photometry \label{tab:k2_phot}}
\tablecolumns{5}
\tablenum{2}
\tablewidth{0pt}
\tablehead{
\colhead{UT Date} &
\colhead{MJD} &
\twocolhead{Phase\tablenotemark{a}} & 
\twocolhead{Time from detection\tablenotemark{b}} & \colhead{Value} \\
\colhead{(YYYY-mm-dd)} & \colhead{(Days)} & \twocolhead{(Rest-frame days)} & \twocolhead{(Rest-frame days)} & \colhead{(Mag)}
}
\decimals
\startdata
2017-12-07.99 & 58094.99 & \phn \phn -66.85 & \phn \phn -48.86 & 20.743$^{c}$ \\
2017-12-08.01 & 58095.01 & -66.83 & -48.84 & 20.744$^{c}$ \\
2017-12-08.05 & 58095.05 & -66.79 & -48.80 & 20.745$^{c}$ \\
2017-12-08.07 & 58095.07 & -66.77 & -48.78 & 20.745$^{c}$ \\
2017-12-08.09 & 58095.09 & -66.75 & -48.76 & 20.745$^{c}$ \\
2017-12-08.11 & 58095.11 & -66.73 & -48.74 & 20.745$^{c}$ \\
\enddata
\tablenotetext{a}{Relative to MJD$_{K2}^{\rm peak} = 58162.1$} \tablenotetext{b}{Relative to MJD$_{\rm detection}^{\rm K2} = 58144.39$} \tablenotetext{c}{3$\sigma$ upper limit}
\tablecomments{The complete K2 light curve is available in the electronic
edition.}
\end{deluxetable*}

\bibliographystyle{aasjournal} 
\bibliography{SN2018oh}

\end{document}